\documentclass{iopart}
\usepackage{graphicx}
\begin{document}
%%%%%%%%%%%%%%%%%%%%%%%%%%%%%%%%%%%%%%%%%%%%%%%%%%%%%%%%%%%%%%%%%%%%%
\title{Max-plus analysis on some binary particle systems}
\author{Daisuke Takahashi$^1$, Junta Matsukidaira$^2$, Hiroaki Hara$^1$, Bao-Feng Feng$^3$}
\address{$^1$ Major in Pure and Applied Mathematics, 3-4-1, Okubo, Shinjuku-ku, Tokyo 169-8555, Japan}
\address{$^2$ Department of Applied Mathematics and Informatics, Ryukoku University, 1-5, Yokotani, Seta Oe-cho, Otsu, Shiga 520-2194, Japan}
\address{$^3$ Department of Mathematics, The University of Texas-Pan American, 1201 West University Drive, Edinburg, TX 78539, USA}
\ead{daisuket@waseda.jp}
\begin{abstract}
We concern with a special class of binary cellular automata, i.e., the so-called particle cellular automata (PCA) in the present paper. We first propose max-plus expressions to PCA of 4 neighbors. Then, by utilizing basic operations of the max-plus algebra and appropriate transformations, PCA4-1, 4-2 and 4-3 are solved exactly and their general solutions are found in terms of max-plus expressions. Finally, we analyze the asymptotic behaviors of general solutions and prove the fundamental diagrams exactly.
\end{abstract}
\submitto{\JPA}
\maketitle
%%%%%%%%%%%%%%%%%%%%%%%%%%%%%%%%%%%%%%%%%%%%%%%%%%%%%%%%%%%%%%%%%%%%%
\section{Introduction}
%%%%%%%%%%%%%%%%%%%%%%%%%%%%%%%%%%%%%%%%%%%%%%%%%%%%%%%%%%%%%%%%%%%%%
  Cellular automaton (CA) can be used as discrete mathematical models for a variety of phenomena\cite{wolfram}. One remarkable feature for CA being used as a model lies in the fact that the setup is mathematically simple. Since both dependent and independent variables are discrete and a set of values which the dependent variable takes is finite, a rule table is more suitable for representing logical expression in a CA model. Thus, CA is free from mathematical conditions such as continuity and differentiability.\par
  Due to the complete discreteness of CA, mathematical tools for analyzing its solutions of CA are very different from those for analyzing solutions of other mathematical objects such as differential equation, finite-difference equation and coupled map lattice. Pattern analysis, graph theory and statistical mathematics are often used to analyze CA \cite{wolfram2}.  However, in seeking for a general solution starting from arbitrary initial data of a CA, this difference often becomes an obstacle. Compact expression of solution using special functions is rarely available for CA due to its discreteness.\par
  One of the bridges between discrete and continuous mathematics to solve this problem was found in the area of integrable system. By applying the following simple formula for the transformation of variables,
\begin{equation}  \label{max-plus}
  \lim_{\varepsilon\to+0}\varepsilon\log(e^{A/\varepsilon}+e^{B/\varepsilon})=\max(A,B),
\end{equation}
we can obtain CA from a finite-difference equation with continuous variables. This procedure is called `ultradiscretization'\cite{tokihiro,matsukidaira}. Solution of the finite-difference equation can be ultradiscretized by the same procedure. For example, we obtain a binary CA called `box and ball system' from the discrete Lotka--Volterra equation through the above transformation\cite{tokihiro,takahashi,takahashi2,mada}.  Moreover, the elementary CA of rule number 184 (ECA184) is obtained from the discrete Burgers equation\cite{nishinari}. ECA184 is considered in the subsequent section.  Note that we can obtain the differential equation and its corresponding solution by taking a continuum limit for a finite-difference equation and its solution.  The continuous analogue to the discrete Lotka--Volterra equation is the Korteweg--de~Vries equation and that to the discrete Burgers equation is the Burgers equation.  Therefore three different discrete levels of equations are directly related each other through two types of limiting procedure.\par
  Replacing basic operations is another way to obtain an ultradiscrete equation from a finite-difference one. Equation (\ref{max-plus}) shows that the addition of the finite-difference equation corresponds to the $\max$ operation of the ultradiscrete one. The multiplication and division correspond to $+$ and $-$ respectively since we have
\begin{equation*}
\fl  \Bigl(\lim_{\varepsilon\to+0}\Bigr)\varepsilon\log(e^{A/\varepsilon}\times e^{B/\varepsilon})=A+B, \qquad
  \Bigl(\lim_{\varepsilon\to+0}\Bigr)\varepsilon\log(e^{A/\varepsilon}/e^{B/\varepsilon})=A-B.
\end{equation*}
Though the subtraction is not well-defined for the ultradiscrete equation, we can automatically obtain the ultradiscrete equation and its solution through a procedure of replacing $+$, $\times$ and $/$ by $\max$, $+$ and $-$ respectively if the subtraction is not included explicitly in the finite-difference equation and its solution.\par
  The algebra defined by operations of $\max$, $+$ and $-$ is called `max-plus algebra'\cite{gaubert}.  Binary operations such as AND, OR and NOT can be easily converted to max-plus operations. Since CA can be defined by binary operations, a rule table of CA is usually transformed into a max-plus equation.  In contrast to a binary solution to the original CA, the solution to a max-plus equation is real-valued. In this sense, both the equation and the solution are extended to the real value range when we use max-plus algebra.  We should point out that it is important but difficult to express a rule table of CA by an appropriate max-plus equation in order to grasp the behavior of solution analytically.\par
  In the present paper, we propose max-plus expression for rule table, general solution and asymptotic behavior of a class of evolutional binary CA.  The sum of values of dependent variables at all spatial sites is conserved through time evolution in this class. ECA184 is included in this class.  Such CA can be used as an important digital model for transportation or flow phenomena. In regard to the asymptotic behavior of solution for this type of CA, a phase transition can often be observed \cite{fuks,fuks2}. In order to clarify the asymptotic behavior mathematically, finding a general solution for this type of CA is necessary. It is found that the max-plus expression is more suitable in finding such a general solution.\par
  The present paper is organized as follows.  In Section~\ref{sec:classification}, we define the class of CA as mentioned above and show examples of solution and observed asymptotic properties.  In Section~\ref{sec:max-plus analysis}, by giving the max-plus expressions of evolution rules, we present general solutions of the CA and, further, analyze the asymptotic behavior based on general solutions.  In Section~\ref{sec:remarks}, concluding remarks are given.  In \ref{sec:formulas}, some convenient formulas about max and min operations are listed.  In \ref{sec:proof solution}, we prove the solution to a specific CA.  In \ref{sec:proof asymptotic}, we prove the asymptotic behavior of the solution.
%%%%%%%%%%%%%%%%%%%%%%%%%%%%%%%%%%%%%%%%%%%%%%%%%%%%%%%%%%%%%%%%%%%%%
\section{Classification of binary particle CA}  \label{sec:classification}
%%%%%%%%%%%%%%%%%%%%%%%%%%%%%%%%%%%%%%%%%%%%%%%%%%%%%%%%%%%%%%%%%%%%%
Let us consider the following evolution equation,
\begin{equation}  \label{gen eq}
  u_j^{n+1}=f(u_{j+r_1}^n, u_{j+r_1+1}^n, \ldots, u_{j+r_2}^n),\qquad \mbox{($-\infty<j<\infty$, $0\le n$)}
\end{equation}
where $n$ and $j$ are integers, representing the time step and the space site number, respectively.  $r_1$ and $r_2$ are also integer constants.  Assume that the value of $u$ takes either 0 or 1, thus  the above equation (\ref{gen eq}) gives an evolution rule of binary CA.  The binary function $f$ has $R$ ($=r_2-r_1+1$) arguments and we call this CA `$R$-neighbors CA'.
%%%%%%%%%%%%%%%%%%%%%%%%%%%%%%%%%%%%%%%%%%%%%%%%%%%%%%%%%%%%%%%%%%%%%
\subsection{Identification of CA}  \label{identification}
%%%%%%%%%%%%%%%%%%%%%%%%%%%%%%%%%%%%%%%%%%%%%%%%%%%%%%%%%%%%%%%%%%%%%
Using a transformation of coordinates defined by $j\to j-cn$, the above equation (\ref{gen eq}) can be rewritten as
\begin{equation}  \label{trans1}
  u_j^{n+1}=f(u_{j+r_1+c}^n, u_{j+r_1+c+1}^n, \ldots, u_{j+r_2+c}^n).
\end{equation}
Thus, the solutions to these two equations (\ref{gen eq}) and (\ref{trans1}) can be transformed from one to the other. Therefore, we can identify equation (\ref{trans1}) for any $c$ with (\ref{gen eq}).\par
  Applying the transformation $j\to -j$ to (\ref{gen eq}), we obtain
\begin{equation}  \label{trans2}
  u_j^{n+1}=f(u_{j-r_2}^n,u_{j-r_2+1}^n,\ldots,u_{j-r_1}^n),
\end{equation}
and can identify (\ref{trans2}) with (\ref{gen eq}) by the same reasoning.  Moreover, if we `flip' the dependent variable $u$, that is, if we apply the transformation $u\to 1-u$, it follows that
\begin{eqnarray}
  u_j^{n+1}&=1-f(1-u_{j+r_1}^n,1-u_{j+r_1+1}^n,\ldots,1-u_{j+r_2}^n) \nonumber \\
  &=g(u_{j+r_1}^n, u_{j+r_1+1}^n, \ldots, u_{j+r_2}^n).  \label{trans3}
\end{eqnarray}
Although the function $g$ is generally different from $f$,  we can still identify (\ref{trans3}) with (\ref{gen eq}).\par
  Moreover, there are cases where $f$ of (\ref{gen eq}) does not depend on all of its arguments, for example, in the following equation
\begin{equation*}
  u_j^{n+1}=f(u_j^n,u_{j+1}^n,u_{j+2}^n), \qquad f(a,b,c)=g(a,b),
\end{equation*}
the number of formal arguments of $f$ is three (3-neighbors) but $f$ does not depend on the last argument. Therefore, this equation is reduced to $u_j^{n+1}=g(u_j^n,u_{j+1}^n)$, which is of 2-neighbors CA.  Let us consider another equation,
\begin{equation*}
  u_j^{n+1}=f(u_j^n,u_{j+1}^n,u_{j+2}^n), \qquad f(a,b,c)=g(a,c),
\end{equation*}
in which, the values of $u$'s at even (odd) sites at time $n+1$ depend only on those at even (odd) sites at time $n$.  Thus, this equation is reduced to $u_j^{n+1}=g(u_j^n,u_{j+1}^n)$, which becomes 2-neighbors CA by the separation of even and odd sites.  In the present paper, we exclude the equations which are reductive to those of less than $R$-neighbors from a class of $R$-neighbors.
%%%%%%%%%%%%%%%%%%%%%%%%%%%%%%%%%%%%%%%%%%%%%%%%%%%%%%%%%%%%%%%%%%%%%
\subsection{Particle CA}
%%%%%%%%%%%%%%%%%%%%%%%%%%%%%%%%%%%%%%%%%%%%%%%%%%%%%%%%%%%%%%%%%%%%%
Let us consider the following condition for any $n$,
\begin{equation}  \label{cons}
  \sum_{j=-\infty}^\infty u_j^{n+1}=\sum_{j=-\infty}^\infty u_j^n,
\end{equation}
which means the sum of $u$, that is, the number of 1's at all space sites is conserved for arbitrary $n$.  Let $u_j^n$ denote the number of particle at $j$-th site and $n$-th time step, and each `1' in the solution represent a particle. Therefore, the particles move among the sites according to the evolution rule defined by (\ref{gen eq}) without creation or annihilation.  We call the CA satisfying the condition (\ref{cons}) `particle CA' (PCA) in this sense.  Classification and statistical analysis of PCA was reported by Fuk\'s and Boccara\cite{fuks,boccara}.\par
  If a CA defined by (\ref{gen eq}) is PCA under appropriate boundary conditions, it can always be written in a conservation form,
\begin{equation}  \label{flux}
\fl  u_j^{n+1}=u_j^n+q(u_{j+r_1}^n,u_{j+r_1+1}^n,\ldots,u_{j+r_2-1}^n)-q(u_{j+r_1+1}^n,u_{j+r_1+2}^n,\ldots,u_{j+r_2}^n)
\end{equation}
where $q(a_1,a_2,\ldots,a_{R-1})$ is a `flux' for the `density' $u$\cite{hattori}. The condition $q(0,0,\ldots,0)=0$ is necessary for a particle CA because a system with zero particle will produce zero flux. This condition can be used to determine an indefinite constant in the expression of $q$.\par
  In what follows, we list up all PCA of 1-neighbor to 4-neighbors identified in Subsection~\ref{identification}.  Label `$n$' or `$n$-$m$' is assigned to each PCA where $n$ denotes the number of neighbors, and $m$ denotes the subclass of the same neighbors CA if it exists. Rules of PCA are defined in the form of (\ref{flux}) together with a rule table of $q$. The upper row of the table shows all possible combinations of binary arguments and the lower row gives the values of
$q$.
\begin{itemize}
\item PCA1: $u_j^{n+1}=u_j^n$, \qquad $q=0$
\item PCA2: none
\item PCA3: $u_j^{n+1}=u_j^n+q(u_{j-1}^n,u_j^n)-q(u_j^n,u_{j+1}^n)$
\begin{equation*}
\begin{array}{|c||c|c|c|c|}
\hline
  a\,b     & 11 & 10 & 01 & 00 \\
\hline
  q(a,b) &  0 &  1 &  0 &  0 \\
\hline
\end{array}
\end{equation*}
\item PCA4: $u_j^{n+1}=u_j^n+q(u_{j-2}^n,u_{j-1}^n,u_j^n)-q(u_{j-1}^n,u_j^n,u_{j+1}^n)$
\begin{equation*}
\fl\begin{array}{ll}
\mbox{4-1} &
\begin{array}{|c||c|c|c|c|c|c|c|c|}
\hline
  a\,b\,c      & 111 & 110 & 101 & 100 & 011 & 010 & 001 & 000 \\
\hline
  q(a,b,c) &  0  &  1  &  0  &  1  &  0  &  1  &  0  &  0  \\
\hline
\end{array}
\medskip
\\
\mbox{4-2} &
\begin{array}{|c||c|c|c|c|c|c|c|c|}
\hline
  a\,b\,c      & 111 & 110 & 101 & 100 & 011 & 010 & 001 & 000 \\
\hline
  q(a,b,c) &  0  &  1  &  0  &  0  &  0  &  0  & -1  &  0  \\
\hline
\end{array}
\medskip
\\
\mbox{4-3} &
\begin{array}{|c||c|c|c|c|c|c|c|c|}
\hline
  a\,b\,c      & 111 & 110 & 101 & 100 & 011 & 010 & 001 & 000 \\
\hline
  q(a,b,c) &  0  &  1  &  0  &  0  &  0  &  0  &  0  &  0  \\
\hline
\end{array}
\medskip
\\
\mbox{4-4} &
\begin{array}{|c||c|c|c|c|c|c|c|c|}
\hline
  a\,b\,c      & 111 & 110 & 101 & 100 & 011 & 010 & 001 & 000 \\
\hline
  q(a,b,c) &  0  &  0  &  0  &  0  &  0  &  1  &  0  &  0  \\
\hline
\end{array}
\end{array}
\end{equation*}
\end{itemize}
For simplicity, let $n=0$ represent the initial time step of the evolution.
For PCA1, we have always a trivial static solution $u_j^n=u_j^0$. PCA3 is equivalent to ECA184 \cite{wolfram}, which is related to a simple traffic congestion model \cite{nishinari}.  In what follows, we will discuss the dynamics of particles and the asymptotic behavior of solutions for PCA3 and 4. Although the max-plus analysis of PCA3 (ECA184) was already given in \cite{nishinari},  we here give a briefly review as a typical scenario of max-plus analysis.\par
  From the rule table of $q$, we can determine the dynamics of a particle system.  In what follows, we show the motion representations of PCA3 and PCA4.  Note that `1' denotes a site occupied by a particle and `0' denotes an empty site. The motion representation for each PCA gives the configurations in which only neighborhoods resulting in a particle motion assuming that by default, in all other cases, particles do not move \cite{fuks2}. The arrow indicates the initial and final positions for a moving particle in one time step.
\begin{equation*}
\begin{array}{ll}
\mbox{PCA3:\quad}  \includegraphics[scale=1.2]{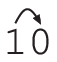} \\
\mbox{PCA4-1:\quad}  \includegraphics[scale=1.2]{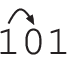}\quad \includegraphics[scale=1.2]{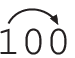}
&\qquad \mbox{PCA4-2:\quad}  \includegraphics[scale=1.2]{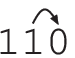}\quad \includegraphics[scale=1.2]{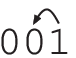} \\
\mbox{PCA4-3:\quad}  \includegraphics[scale=1.2]{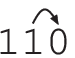}&\qquad \mbox{PCA4-4:\quad}  \includegraphics[scale=1.2]{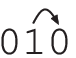}
\end{array}
\end{equation*}
%%%%%%%%%%%%%%%%%%%%%%%%%%%%%%%%%%%%%%%%%%%%%%%%%%%%%%%%%%%%%%%%%%%%%
\subsection{Asymptotic behavior of PCA}
%%%%%%%%%%%%%%%%%%%%%%%%%%%%%%%%%%%%%%%%%%%%%%%%%%%%%%%%%%%%%%%%%%%%%
In this subsection, we discuss the asymptotic behavior of solution from arbitrary initial data as $n\to\infty$ . The behavior is mainly dependent on the density of particles, or the average number of particles per site. To observe this dependence, a periodic boundary condition for space sites is assumed. Let $K$ be the number of sites in one period, the density $\rho$ is defined by
\begin{equation*}
  \rho=\frac{1}{K}\sum_{j=1}^K u_j^n.
\end{equation*}
Since the number of particles is conserved, $\rho$ is also a constant irrespective of time $n$. \par
\begin{figure}[p]
\begin{tabular}{c}
\includegraphics[scale=0.55]{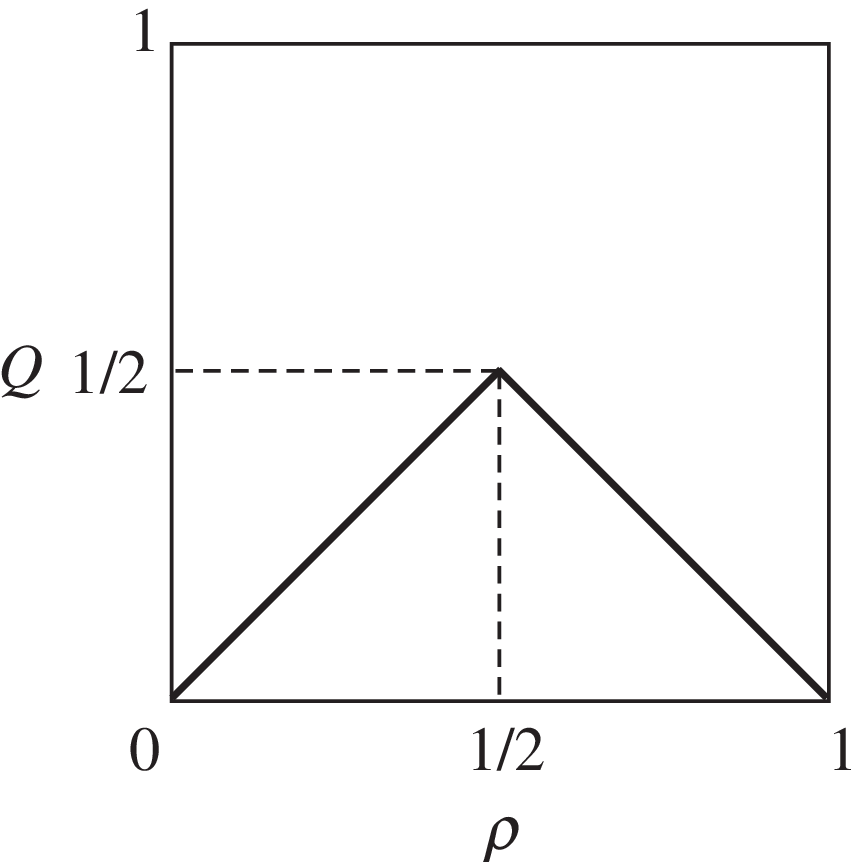}
\\
\begin{tabular}{c}
\includegraphics[scale=0.9]{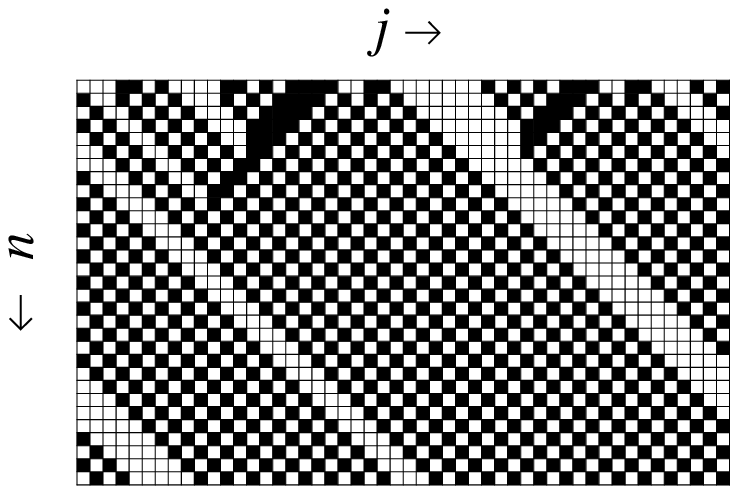} \\
$\rho=0.45$
\end{tabular}
\\
\begin{tabular}{c}
\includegraphics[scale=0.9]{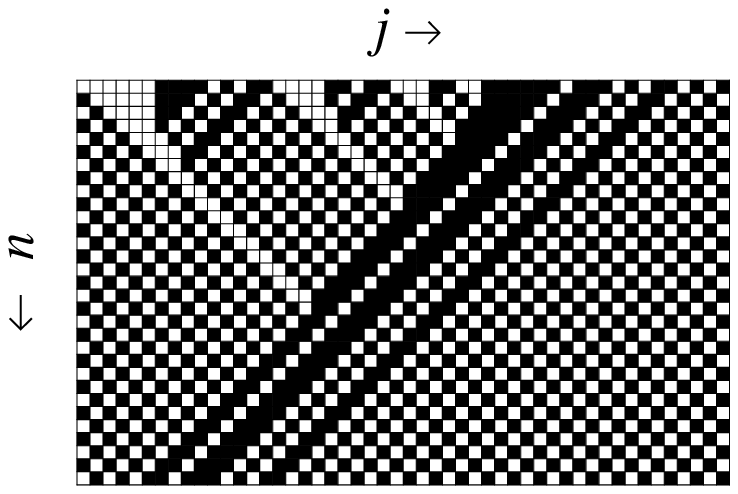} \\
$\rho=0.55$
\end{tabular}
\\
\end{tabular}
\caption{FD and solutions of PCA3}
\label{fig:PCA3}
\end{figure}
\begin{figure}[p]
\begin{tabular}{c}
\includegraphics[scale=0.55]{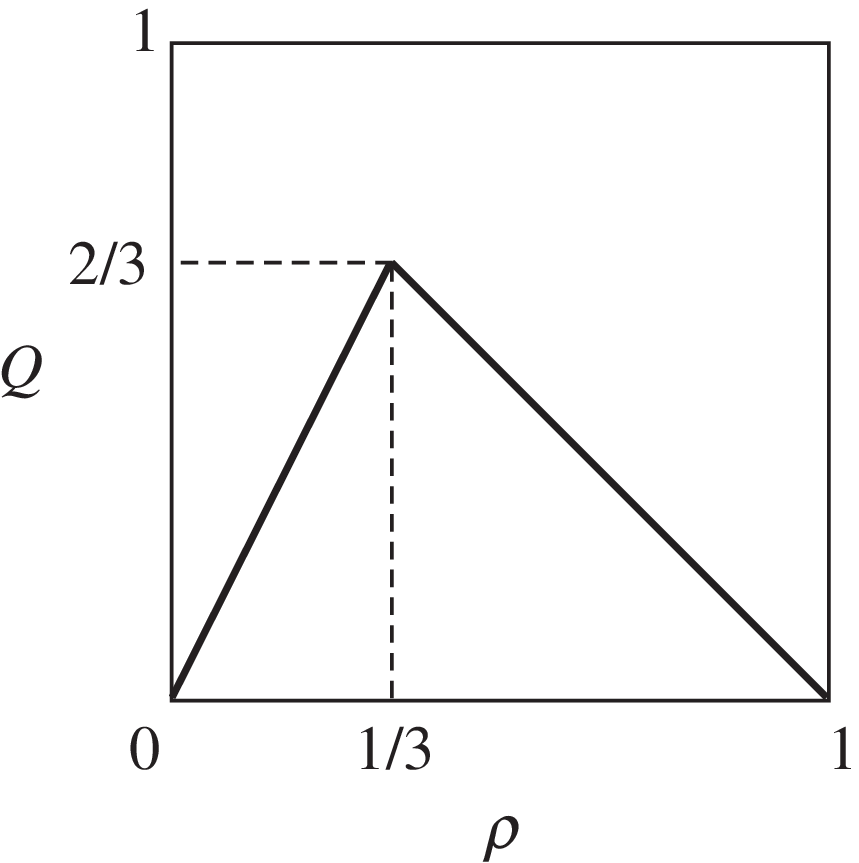}
\\
\begin{tabular}{c}
\includegraphics[scale=0.9]{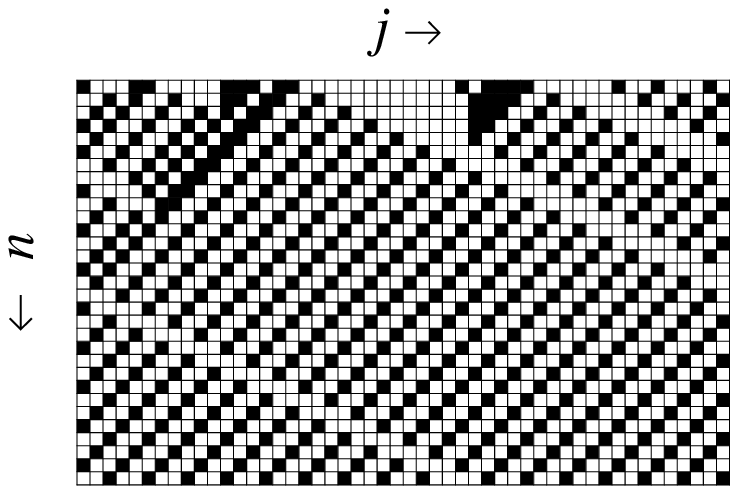} \\
$\rho=0.32$
\end{tabular}
\\
\begin{tabular}{c}
\includegraphics[scale=0.9]{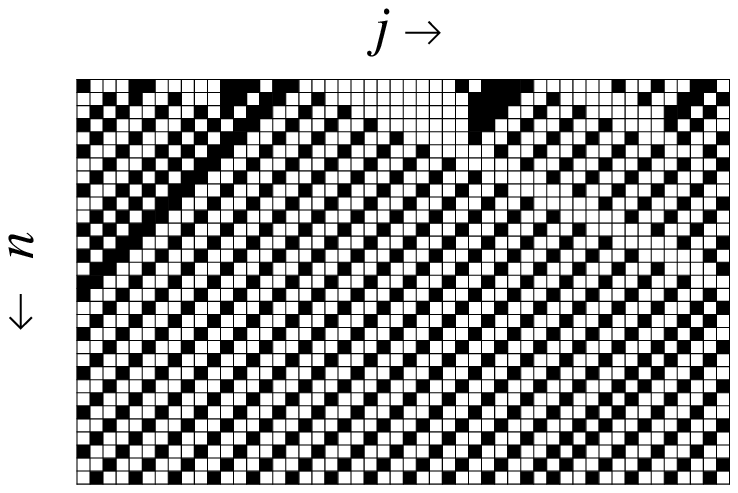} \\
$\rho=0.34$
\end{tabular}
\\
\begin{tabular}{c}
\includegraphics[scale=0.9]{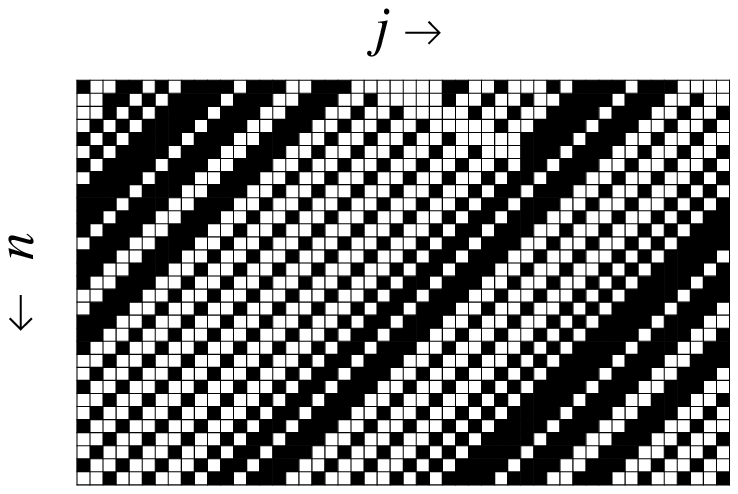} \\
$\rho=0.5$
\end{tabular}
\\
\end{tabular}
\caption{FD and solutions of PCA4-1}
\label{fig:PCA4-1}
\end{figure}
\begin{figure}[p]
\begin{tabular}{c}
\includegraphics[scale=0.55]{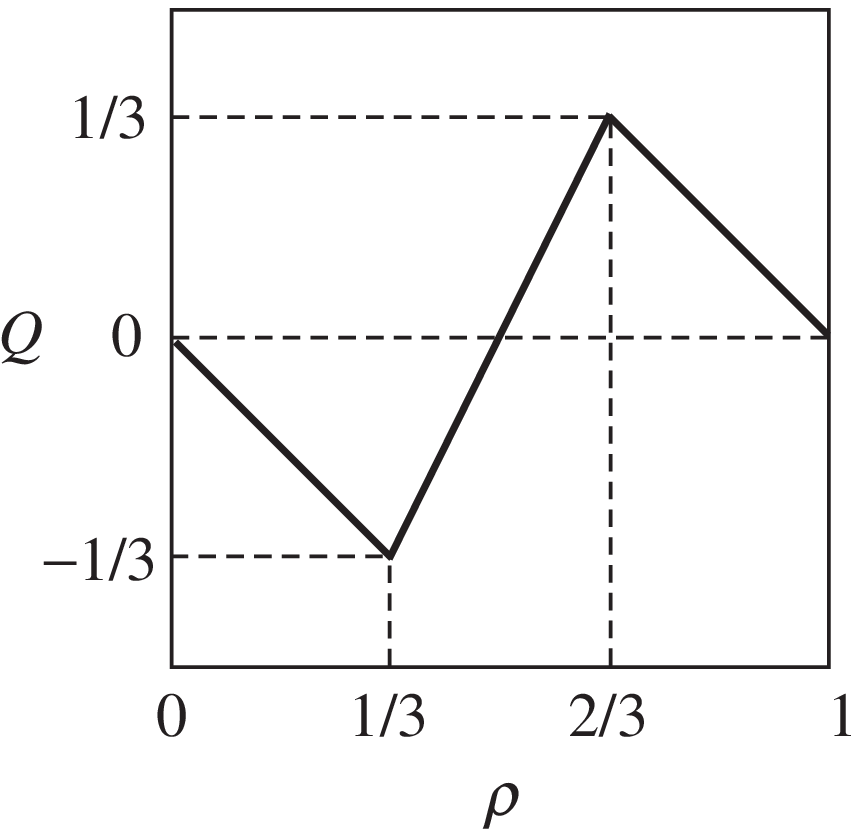}
\\
\begin{tabular}{c}
\includegraphics[scale=0.9]{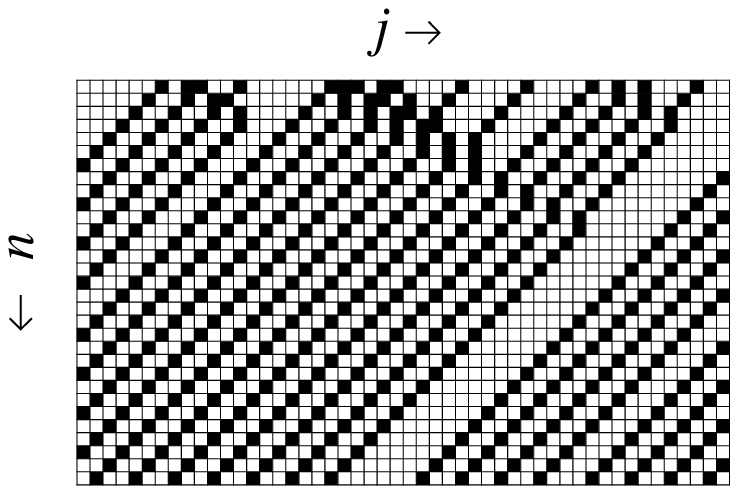} \\
$\rho=0.3$
\end{tabular}
\\
\begin{tabular}{c}
\includegraphics[scale=0.9]{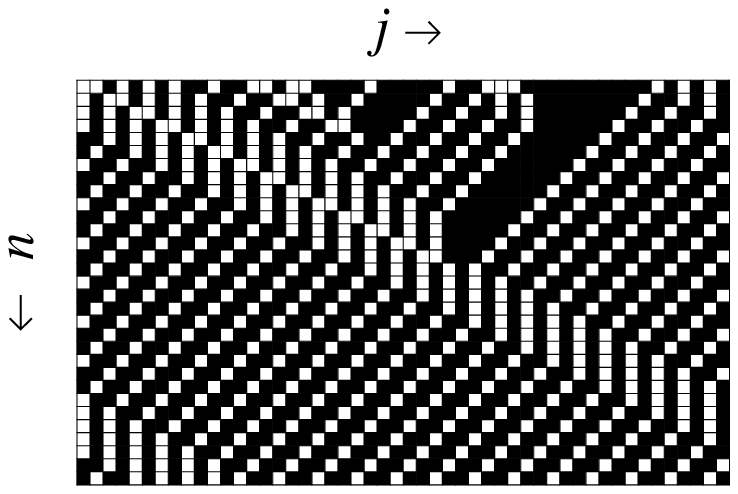} \\
$\rho=0.64$
\end{tabular}
\\
\begin{tabular}{c}
\includegraphics[scale=0.9]{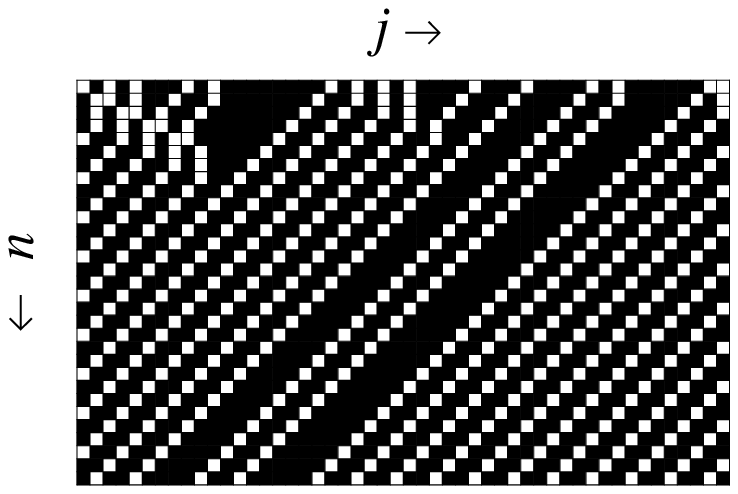} \\
$\rho=0.7$
\end{tabular}
\\
\end{tabular}
\caption{FD and solutions of PCA4-2}
\label{fig:PCA4-2}
\end{figure}
\begin{figure}[p]
\begin{tabular}{c}
\includegraphics[scale=0.55]{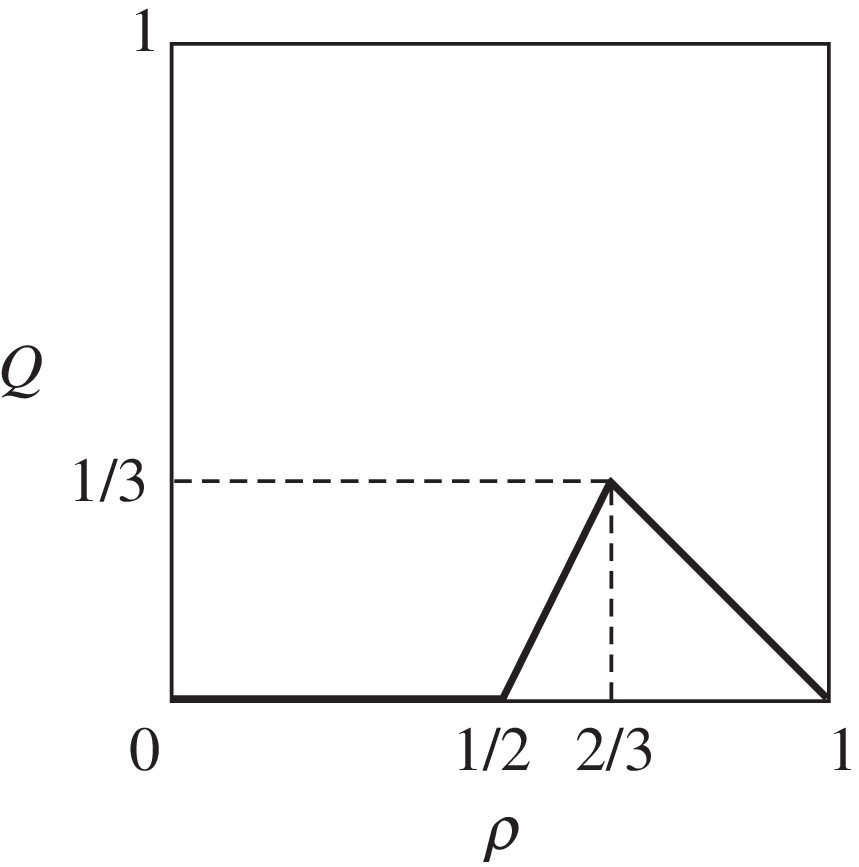}
\\
\begin{tabular}{c}
\includegraphics[scale=0.9]{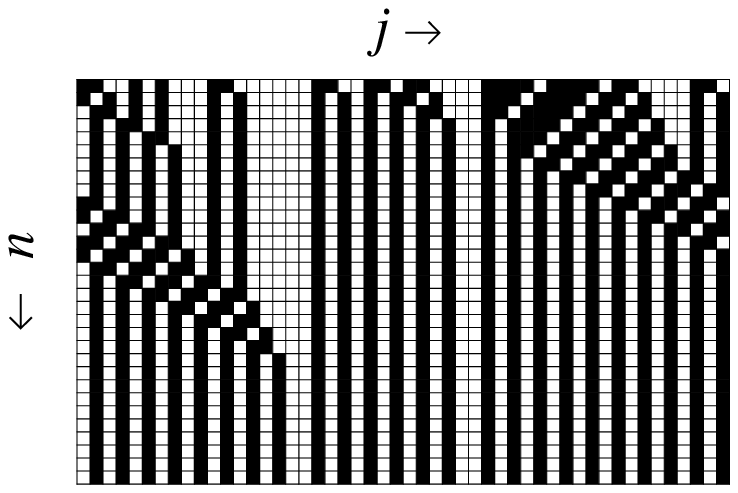} \\
$\rho=0.48$
\end{tabular}
\\
\begin{tabular}{c}
\includegraphics[scale=0.9]{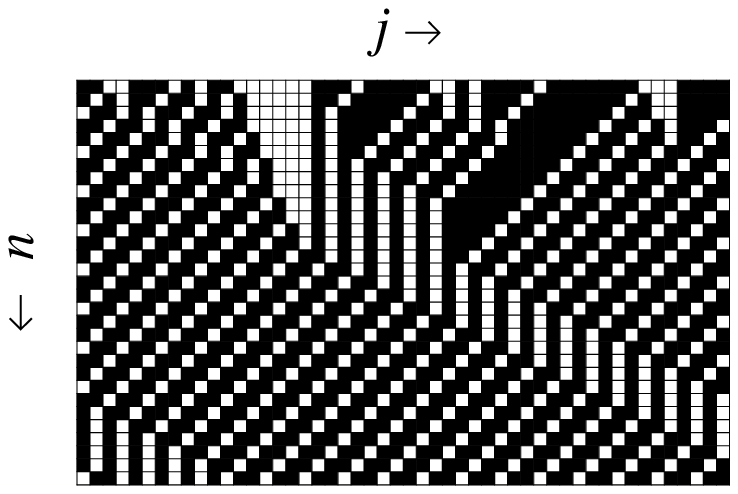} \\
$\rho=0.64$
\end{tabular}
\\
\begin{tabular}{c}
\includegraphics[scale=0.9]{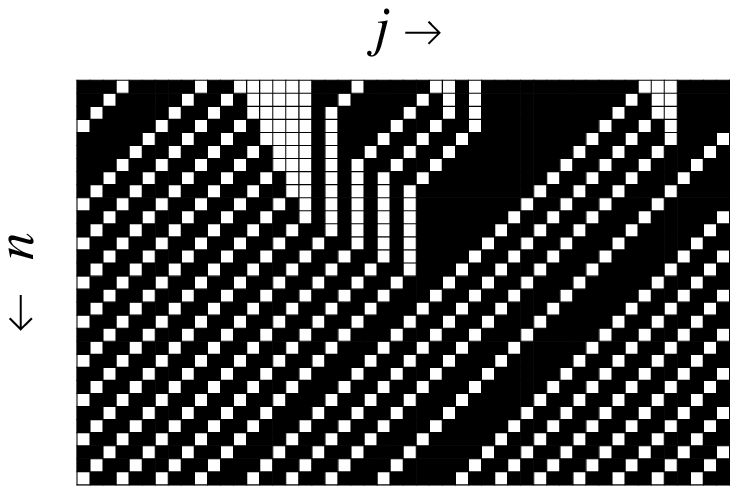} \\
$\rho=0.7$
\end{tabular}
\\
\end{tabular}
\caption{FD and solutions of PCA4-3}
\label{fig:PCA4-3}
\end{figure}
\begin{figure}[p]
\begin{tabular}{c}
\includegraphics[scale=0.6]{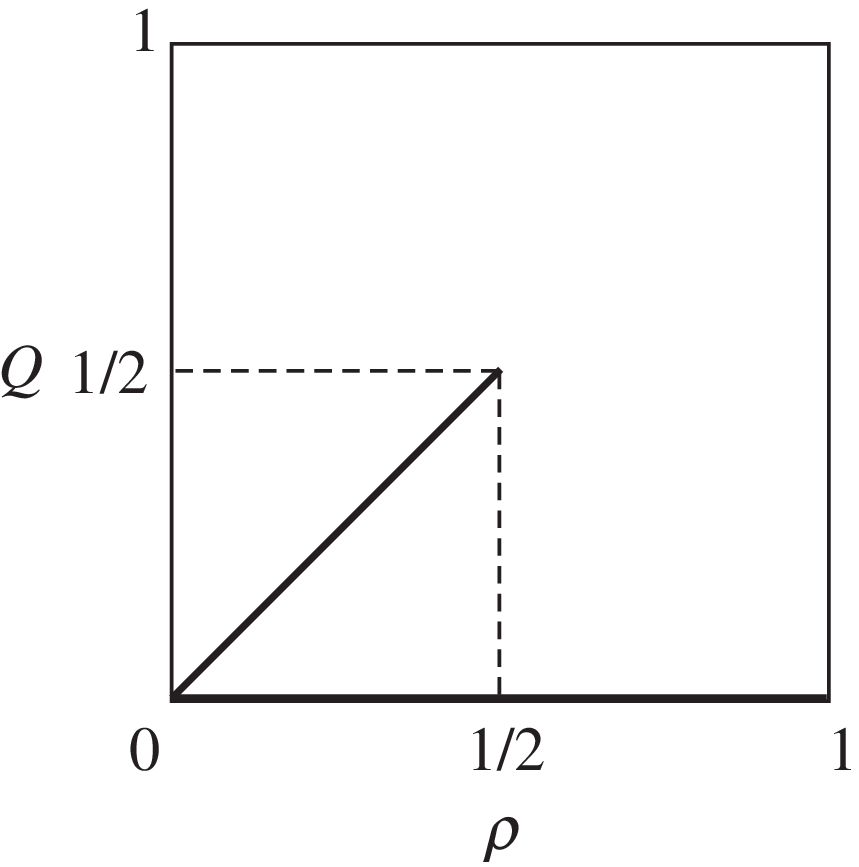}
\\
\begin{tabular}{c}
\includegraphics[scale=0.9]{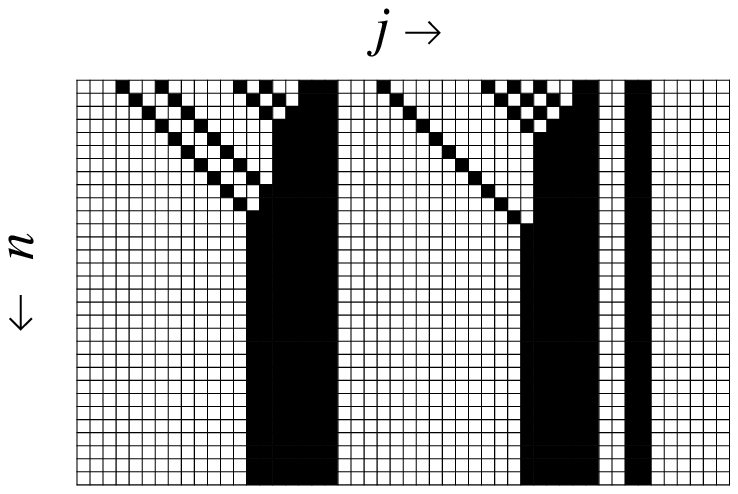} \\
$\rho=0.3$
\end{tabular}
\\
\begin{tabular}{c}
\includegraphics[scale=0.9]{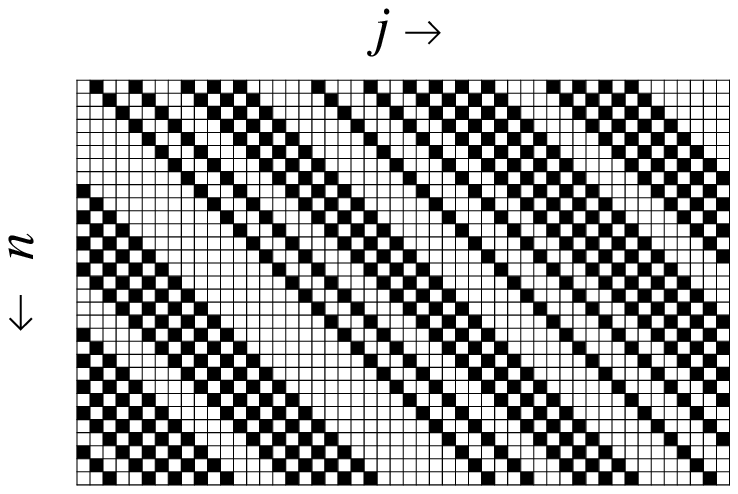} \\
$\rho=0.3$
\end{tabular}
\\
\begin{tabular}{c}
\includegraphics[scale=0.9]{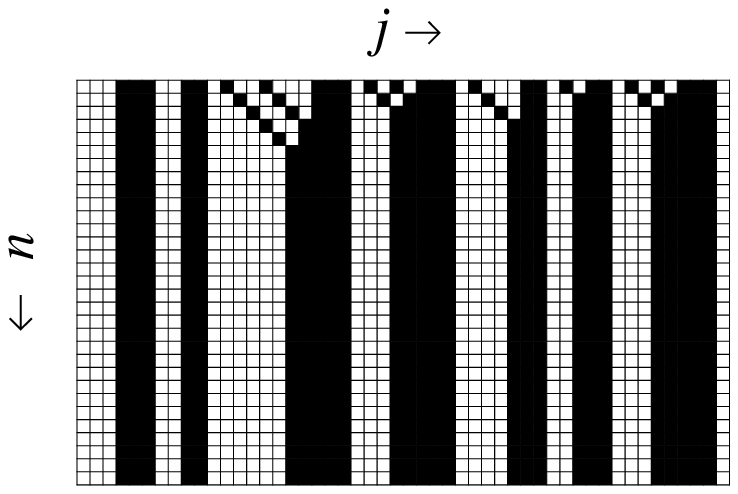} \\
$\rho=0.52$
\end{tabular}
\\
\end{tabular}
\caption{FD and solutions of PCA4-4}
\label{fig:PCA4-4}
\end{figure}

  For most of PCA, the asymptotic behavior changes at certain critical values of $\rho$. The relation between the density and the average flux per site reflects this critical phenomenon well.  The average flux $\overline{q}^n$ per site is defined by
\begin{equation*}
  \overline{q}^n=\frac{1}{K}\sum_{j=1}^K q(u_j^n,\ldots,u_{j+R-1}^n).
\end{equation*}
For large enough $n$, the evolution of a PCA often approaches a steady state, in which $\overline{q}^n$ becomes a constant. In the case of convergence, the constant is defined by
\begin{equation*}
  Q=\lim_{n\to\infty}\overline{q}^n.
\end{equation*}
The value of $Q$ is independent of the initial distribution of particles and depends solely on the density $\rho$ for PCA3, 4-1, 4-2 and 4-3, whereas for PCA4-4, the value of $Q$ depends on both the density $\rho$ and the initial condition. The graph of $Q$ vs. $\rho$ is called `fundamental diagram' in the area of traffic flow analysis.  We use this terminology throughout the present paper and abbreviate it as FD. The FD's for PCA3 and PCA4, as well as examples of evolutions are shown in Figures~\ref{fig:PCA3}-\ref{fig:PCA4-4} respectively.\par
%%%%%%%%%%%%%%%%%%%%%%%%%%%%%%%%%%%%%%%%%%%%%
%%%%%%%%%%%%%%%%%%%%%%%%%%%%%%%%%%%%%%%%%%%%%
%%%%%%%%%%%%%%%%%%%%%%%%%%%%%%%%%%%%%%%%%%%%%
%%%%%%%%%%%%%%%%%%%%%%%%%%%%%%%%%%%%%%%%%%%%%
%%%%%%%%%%%%%%%%%%%%%%%%%%%%%%%%%%%%%%%%%%%%%
%%%%%%%%%%%%%%%%%%%%%%%%%%%%%%%%%%%%%%%%%%%%%
The relations between the solutions of steady states at large enough $n$ and the critical values of $\rho$ are listed below.
\begin{equation*}
\mbox{PCA3:\quad}
  u_j^{n+1}=\cases{u_{j-1}^n  & ($0\le\rho\le1/2$) \\ u_{j+1}^n & ($1/2\le\rho\le1$)}
\end{equation*}
\begin{equation*}
\mbox{PCA4-1:\quad}
  u_j^{n+1}=\cases{u_{j-2}^n & ($0\le\rho\le1/3$) \\ u_{j+1}^n & ($1/3\le\rho\le1$)}
\end{equation*}
\begin{equation*}
\mbox{PCA4-2:\quad}
  u_j^{n+1}=\cases{u_{j+1}^n & ($0\le\rho\le1/3$) \\ u_{j-2}^n & ($1/3\le\rho\le2/3$) \\ u_{j+1}^n & ($2/3\le\rho\le1$)}
\end{equation*}
\begin{equation*}
\mbox{PCA4-3:\quad}
  u_j^{n+1}=\cases{u_j^n & ($0\le\rho\le1/2$) \\ u_{j-2}^n & ($1/2\le\rho\le2/3$) \\ u_{j+1}^n & ($2/3\le\rho\le1$)}
\end{equation*}
\begin{equation*}
\mbox{PCA4-4:\quad}
  u_j^{n+1}=\cases{u_j^n \mbox{\ or\ } u_{j-1}^n & ($0\le\rho\le1/2$) \\ u_j^n & ($1/2\le\rho\le1$)}
\end{equation*}
%%%%%%%%%%%%%%%%%%%%%%%%%%%%%%%%%%%%%%%%%%%%%%%%%%%%%%%%%%%%%%%%%%%%%
\section{Max-plus analysis of PCA}  \label{sec:max-plus analysis}
%%%%%%%%%%%%%%%%%%%%%%%%%%%%%%%%%%%%%%%%%%%%%%%%%%%%%%%%%%%%%%%%%%%%%
\subsection{PCA3 \cite{nishinari}}
%%%%%%%%%%%%%%%%%%%%%%%%%%%%%%%%%%%%%%%%%%%%%%%%%%%%%%%%%%%%%%%%%%%%%
The flux function of PCA3 was found to be $q(a,b)=\min(a,1-b)$ if $a$ and $b$ are binary.  Thus, the evolution rule can be rewritten in the following max-plus form,
\begin{equation}  \label{eq PCA3}
  u_j^{n+1}=u_j^n+\min(u_{j-1}^n,1-u_j^n)-\min(u_j^n,1-u_{j+1}^n).
\end{equation}
Using a transformation of variable from $u$ to $f$,
\begin{equation}  \label{trans PCA3}
  u_j^n=f_j^n-f_{j-1}^n+\frac{1}{2},
\end{equation}
we obtain the evolution rule for $f$
\begin{equation}  \label{potential PCA3}
  f_j^{n+1}=\max(f_{j-1}^n,f_{j+1}^n).
\end{equation}
Note that, for simplicity, an arbitrary constant occurring in the process of deriving (\ref{potential PCA3}) is set to 0.\par
  The initial value problem of (\ref{potential PCA3}) can be easily solved as
\begin{equation*}
  f_j^n = \max(f_{j-n}^0,f_{j-n+2}^0,\ldots,f_{j+n-2}^0,f_{j+n}^0).
\end{equation*}
The inverse relation of (\ref{trans PCA3}) is
\begin{equation*}
  f_j^n=\sum_{k=0}^j\Bigl(u_k^n-\frac{1}{2}\Bigr) + c,
\end{equation*}
with an arbitrary constant $c$, from which we have
\begin{equation*}
  f_{j+K}^n-f_j^n=\sum_{k=j+1}^{j+K}\Bigl(u_k^n-\frac{1}{2}\Bigr)=K\Bigl(\rho-\frac{1}{2}\Bigr).
\end{equation*}
Therefore, it is obvious that $f_{j+K}^n\le f_j^n$ for $\rho\le1/2$ and $f_{j+K}^n\ge f_j^n$ for $\rho\ge1/2$.  Using above inequalities, we have
\begin{equation*}
  f_j^n
  =\cases{
    \max(f_{j-n}^0,f_{j-n+2}^0,\ldots,f_{j-n+2K}^0) & ($0\le\rho\le1/2$) \\
    \max(f_{j+n-2K}^0,f_{j+n-2K+2}^0,\ldots,f_{j+n}) & ($1/2\le\rho\le1$)
  }
\end{equation*}
for large enough $n$. Furthermore, we conclude that
\begin{equation*}
\left\{
\begin{array}{lll}
  f_j^{n+1}=f_{j-1}^n, & u_j^{n+1}=u_{j-1}^n & (0\le\rho\le1/2) \\
  f_j^{n+1}=f_{j+1}^n, & u_j^{n+1}=u_{j+1}^n & (1/2\le\rho\le1)
\end{array}
\right.
\end{equation*}
for large enough $n$.  These analytical relations explain the steady solution of PCA3 shown in Figure~\ref{fig:PCA3}.\par
  Finally, substituting the relation $u_j^{n+1}=u_{j-1}^n$ under the case of $0\le\rho\le1/2$ into the evolution equation (\ref{eq PCA3}) , we arrive at
\begin{equation*}
  u_{j-1}^n-\min(u_{j-1}^n,1-u_j^n)=u_j^n-\min(u_j^n,1-u_{j+1}^n)
\end{equation*}
for any $j$. In other words, $u_{j-1}^n-\min(u_{j-1}^n,1-u_j^n)$ is a constant irrespective of $j$.  Since $0\le\rho\le1/2$, there exists at least one site $j$ satisfying $(u_{j-1}^n,u_j^n)=(0,0)$ or $(1,0)$, which resulting in a zero value for the above constant. Thus, we have
\begin{equation*}
  Q=\frac{1}{K}\sum_{j=1}^K\min(u_{j-1}^n,1-u_j^n)=\frac{1}{K}\sum_{j=1}^K u_{j-1}^n=\rho
\end{equation*}
in the case of $0\le\rho\le1/2$.  Similarly we obtain $Q=1-\rho$ in the case of $1/2\le\rho\le1$.  This relation gives FD of PCA3 as shown in Figure~\ref{fig:PCA3}.
%%%%%%%%%%%%%%%%%%%%%%%%%%%%%%%%%%%%%%%%%%%%%%%%%%%%%%%%%%%%%%%%%%%%%
\subsection{PCA4-1}
%%%%%%%%%%%%%%%%%%%%%%%%%%%%%%%%%%%%%%%%%%%%%%%%%%%%%%%%%%%%%%%%%%%%%
This PCA is known as the Fukui--Ishibashi model used for traffic congestion analysis\cite{fukui} and its Euler and Lagrange representation are studied using max-plus expression\cite{matsukidaira2}.  Asymptotic behavior for the solution of PCA4-1 is similar to that of PCA3 except a few slight differences: \romannumeral1) critical value of $\rho$ is $1/3$, \romannumeral2) for the smaller region of $\rho$ ($0\le\rho\le1/3$), all particles move with a speed of 2 at large enough $n$.  The evolution equation of PCA4-1 can be described in the following max-plus form,
\begin{equation*}
  u_j^{n+1}=u_j^n+\min(u_{j-2}^n+u_{j-1}^n,1-u_j^n)-\min(u_{j-1}^n+u_j^n,1-u_{j+1}^n).
\end{equation*}
Introducing a transformation
\begin{equation}  \label{trans PCA4-1}
  u_j^n=f_j^n-f_{j-1}^n+\frac{1}{3},
\end{equation}
we obtain
\begin{equation*}
  f_j^{n+1}=\max(f_{j-2}^n,f_{j+1}^n)
\end{equation*}
by choosing an appropriate value for the arbitrary constant.\par
  Consequently, the general solution for the initial value problem turns out to be
\begin{equation*}
  f_j^n=\max(f_{j-2n}^0,f_{j-2n+3}^0,\ldots,f_{j+n-3}^0,f_{j+n}^0).
\end{equation*}
On the other hand, the inverse relation of (\ref{trans PCA4-1}) gives
\begin{equation*}
  f_j^n=\sum_{k=0}^j\Bigl(u_k^n-\frac{1}{3}\Bigr)+c,
\end{equation*}
which leads to
\begin{equation*}
  f_{j+K}^n-f_j^n=K\Bigl(\rho-\frac{1}{3}\Bigr).
\end{equation*}
By similar calculations for PCA3, at large enough $n$, we can derive steady state solutions:
\begin{equation*}
\left\{
\begin{array}{lll}
  f_j^{n+1}=f_{j-2}^n, & u_j^{n+1}=u_{j-2}^n & (0\le\rho\le1/3) \\
  f_j^{n+1}=f_{j+1}^n, & u_j^{n+1}=u_{j+1}^n & (1/3\le\rho\le1)
\end{array}
\right.,
\end{equation*}
and further the following relations
\begin{equation*}
\min(u_{j-1}^n+u_j^n,1-u_{j+1}^n)
= \cases{
  u_{j-1}^n+u_j^n & ($0\le\rho\le1/3$) \\
  1-u_{j+1}^n & ($1/3\le\rho\le1$)
}.
\end{equation*}
From above relations, it can be easily shown that
\begin{equation}
  Q=\cases{2\rho & ($0\le\rho\le1/3$) \\ 1-\rho & ($1/3\le\rho\le1$)}.
\end{equation}
In summary, the above analytical results agree with the solutions and FD shown in Figure~\ref{fig:PCA4-1}.
%%%%%%%%%%%%%%%%%%%%%%%%%%%%%%%%%%%%%%%%%%%%%%%%%%%%%%%%%%%%%%%%%%%%%
\subsection{PCA4-2}
%%%%%%%%%%%%%%%%%%%%%%%%%%%%%%%%%%%%%%%%%%%%%%%%%%%%%%%%%%%%%%%%%%%%%
 It can be shown that the evolution equation of PCA4-2 is
\begin{equation}  \label{eq PCA4-2}
\cases{
  u_j^{n+1}=u_j^n+q(u_{j-2}^n,u_{j-1}^n,u_j^n)-q(u_{j-1}^n,u_j^n,u_{j+1}^n)\\
  q(a,b,c)=\min(\max(-c,a+b-1), 1-c)
}.
\end{equation}
Through a transformation,
\begin{equation}  \label{trans PCA4-2}
  u_j^n=f_j^n-f_{j-1}^n+\frac{1}{2},
\end{equation}
it follows, from (\ref{eq PCA4-2}), that the evolution equation to $f$ is
\begin{equation}  \label{potential PCA4-2}
  f_j^{n+1}=\max\Bigl(\min\Bigl(f_{j-2}^n,f_{j+1}^n+\frac{1}{2}\Bigr),f_{j+1}^n-\frac{1}{2}\Bigr).
\end{equation}
The general solution to equation (\ref{potential PCA4-2}) turns out to be
\begin{equation}  \label{sol PCA4-2}
  f_j^n=\max_{0\le k\le n}\Bigl(\min\Bigl(f_{j-2n+3k}^0-\frac{k}{2},\min_{k+1\le i\le n}f_{j-2n+3i}^0+\frac{i}{2}\Bigr)\Bigr).
\end{equation}
\par
  By expanding RHS of (\ref{sol PCA4-2}), specific examples of $f_j^n$ for $n=1$, 2, 3 are given as
\begin{eqnarray*}
\fl f_j^1=\max\Bigl(\min\Bigl(f_{j-2}^0,f_{j+1}^0+\frac{1}{2}\Bigr),f_{j+1}^0-\frac{1}{2}\Bigr), \\
\fl f_j^2=\max\Bigl(\min\Bigl(f_{j-4}^0,f_{j-1}^0+\frac{1}{2},f_{j+2}^0+1\Bigr),\min\Bigl(f_{j-1}^0-\frac{1}{2},f_{j+2}^0+1\Bigr),f_{j+2}^0-1\Bigr), \\
\fl \eqalign{f_j^3=\max\Bigl(&\min\Bigl(f_{j-6}^0,f_{j-3}^0+\frac{1}{2},f_j^0+1,f_{j+3}^{0}+\frac{3}{2}\Bigr),\min\Bigl(f_{j-3}^0-\frac{1}{2},f_j^0+1,f_{j+3}^{0}+\frac{3}{2}\Bigr), \\
    &\min(f_j^0-1,f_{j+3}^{0}+\frac{3}{2}),
    f_{j+3}^0-\frac{3}{2}\Bigr).}
\end{eqnarray*}
The proof for the general solution (\ref{sol PCA4-2}) satisfying (\ref{potential PCA4-2})
is given in \ref{sec:proof solution}.\par
  Moreover, at large enough $n$, the asymptotic behavior of the solution, as well as the value of $Q$, can be shown to be:
\begin{eqnarray}
\fl\eqalign{
f_j^n
=\cases{
  \min_{0\le k\le K-1}\Bigl(f_{j+n-3k}^0+\frac{n-k}{2}\Bigr) \qquad (0\le\rho\le1/3) \\
  \eqalign{\max\Bigl(&\max_{0\le k\le K-2}\Bigl(\min\Bigl(f_{j-2n+3k}^0-\frac{k}{2},\min_{1\le i\le K-1}f_{j-2n+3(k+i)}^0+\frac{k+i}{2}\Bigr)\Bigr),\\
  &\max_{K-1\le k\le 2(K-1)}f_{j-2n+3k}^0-\frac{k}{2}\Bigr) \qquad (1/3\le\rho\le2/3)} \\
  \max_{0\le k\le K-1}\Bigl(f_{j+n-3k}^0-\frac{n-k}{2}\Bigr) \qquad (2/3\le\rho\le1)
}}, \label{asymp PCA4-2 f} \\
\fl\cases{
  f_j^{n+1}=f_{j+1}^n+\frac{1}{2}, \qquad u_j^{n+1}=u_{j+1}^n & ($0\le\rho\le1/3$) \\
  f_j^{n+1}=f_{j-2}^n, \qquad u_j^{n+1}=u_{j-2}^n & ($1/3\le\rho\le2/3$) \\
  f_j^{n+1}=f_{j+1}^n-\frac{1}{2}, \qquad u_j^{n+1}=u_{j+1}^n & ($2/3\le\rho\le1$) \\
}, \label{asymp PCA4-2 u} \\
\fl q(u_{j-2}^n,u_{j-1}^n,u_j^n)
=\cases{
  -u_j^n & ($0\le\rho\le1/3$) \\
  u_{j-2}^n+u_{j-1}^n-1 & ($1/3\le\rho\le2/3$) \\
  1-u_j^n & ($2/3\le\rho\le1$)
}, \label{asymp PCA4-2 q} \\
\fl Q
=\cases{
  -\rho & ($0\le\rho\le1/3$) \\
  2\rho-1 & ($1/3\le\rho\le2/3$) \\
  1-\rho & ($2/3\le\rho\le1$)
}. \label{asymp PCA4-2 Q}
\end{eqnarray}
The proof of the above results is given in \ref{sec:proof asymptotic}. In summary, the analytical results reveal the steady state solutions and FD as shown in Figure~\ref{fig:PCA4-2}.
%%%%%%%%%%%%%%%%%%%%%%%%%%%%%%%%%%%%%%%%%%%%%%%%%%%%%%%%%%%%%%%%%%%%%
\subsection{PCA4-3}
%%%%%%%%%%%%%%%%%%%%%%%%%%%%%%%%%%%%%%%%%%%%%%%%%%%%%%%%%%%%%%%%%%%%%
The evolution equation of PCA4-3 can be expressed by
\begin{equation*}
\label{PCA42-u}
\cases{
  u_j^{n+1}=u_j^n+q(u_{j-2}^n,u_{j-1}^n,u_j^n)-q(u_{j-1}^n,u_j^n,u_{j+1}^n)\\
  q(a,b,c)=\min(\max(0,a+b-1), 1-c)
}.
\end{equation*}
By using a transformation,
\begin{equation*}
  u_j^n=f_j^n-f_{j-1}^n+\frac{1}{2},
\end{equation*}
we arrive at an evolution equation for $f$,
\begin{equation}
\label{PCA42-f}
  f_j^{n+1}=\max\Bigl(\min(f_{j-2}^n,f_j^n),f_{j+1}^n-\frac{1}{2}\Bigr).
\end{equation}
From which, the general solution turns out to be
\begin{equation*}
  f_j^n=\max_{0\le k\le n}\Bigl(\min\Bigl(f_{j-2n+3k}^0-\frac{k}{2},\min_{3k/2<i\le n}f_{j-2n+2i}^0\Bigr)\Bigr)
\end{equation*}
by means of a similar proof as PCA4-2. Therefore, it is omitted here.\par
  The specific expressions of $f_j^n$ for $n=1$, 2, 3 are as follows:
\begin{eqnarray*}
\fl f_j^1=\max\Bigl(\min(f_{j-2}^0,f_j^0),f_{j+1}^0-\frac{1}{2}\Bigr), \\
\fl f_j^2=\max\Bigl(\min(f_{j-4}^0,f_{j-2}^0,f_j^0),\min\Bigl(f_{j-1}^0-\frac{1}{2},f_j^0\Bigr),f_{j+2}^0-1\Bigr), \\
\fl f_j^3=\max\Bigl(\min(f_{j-6}^0,f_{j-4}^0,f_{j-2}^0,f_j^0),\min\Bigl(f_{j-3}^0-\frac{1}{2},f_{j-2}^0,f_j^0\Bigr), f_j^0-1,f_{j+3}^0-\frac{3}{2}\Bigr).
\end{eqnarray*}
\par
At large enough $n$, we can derive the asymptotic behavior of solution and, further, the value of $Q$, which both depend on the density $\rho$ as follows
\begin{equation*}
\fl\eqalign{
f_j^n
=\cases{
  \min_{0\le k\le K-1}f_{j-2k}^0  & ($0\le\rho\le1/2$) \\
  \eqalign{\max_{0\le k\le K-1}\Bigl(&\min\Bigl(f_{j-2n+3k}^0-\frac{k}{2}, \\
  &\min_{1\le i\le K-1}f_{j-2n+2\lfloor3k/2\rfloor+2i}^0\Bigr)\Bigr)} & ($1/2\le\rho\le2/3$) \\
  \max_{0\le k\le K-1}\Bigl(f_{j+n-3k}^0-\frac{n-k}{2}\Bigr) & ($2/3\le\rho\le1$)
}, \\
\cases{
  f_j^{n+1}=f_j^n, \qquad u_j^{n+1}=u_j^n & ($0\le\rho\le1/2$) \\
  f_j^{n+1}=f_{j-2}^n, \qquad u_j^{n+1}=u_{j-2}^n & ($1/2\le\rho\le2/3$) \\
  f_j^{n+1}=f_{j+1}^n-\frac{1}{2}, \qquad u_j^{n+1}=u_{j+1}^n & ($2/3\le\rho\le1$) \\
}, \\
q(u_{j-2}^n,u_{j-1}^n,u_j^n)
=\cases{
  0 & ($0\le\rho\le1/2$) \\
  u_{j-2}^n+u_{j-1}^n-1 & ($1/2\le\rho\le2/3$) \\
  1-u_j^n & ($2/3\le\rho\le1$)
}, \\
Q
=\cases{
  0 & ($0\le\rho\le1/3$) \\
  2\rho-1 & ($1/3\le\rho\le2/3$) \\
  1-\rho & ($2/3\le\rho\le1$)
}.
}
\end{equation*}
Again, the above results reveal the solutions and FD as shown in
Figure~\ref{fig:PCA4-3}. Since the proof is tedious and similar to that of PCA4-2, it is omitted here.
%%%%%%%%%%%%%%%%%%%%%%%%%%%%%%%%%%%%%%%%%%%%%%%%%%%%%%%%%%%%%%%%%%%%%
\subsection{PCA4-4}
%%%%%%%%%%%%%%%%%%%%%%%%%%%%%%%%%%%%%%%%%%%%%%%%%%%%%%%%%%%%%%%%%%%%%
We are unable to find a general solution to this type of PCA in the max-plus form. However, it is worth pointing out the following observations from the binary rule table in the case of $0\le\rho\le1/2$: if there exists a site $j$ satisfying $u_{j-1}^0=u_j^0=1$ in the initial data, the asymptotic solution will become static, that is, $u_j^{n+1}=u_j^n$ at large enough $n$.  If there is no two consecutive 1's in the initial data, the solution $u_j^n$ satisfies $u_{j-1}^n=u_j^n$ for any $j$ and $n$.  It then follows that $Q=0$ in the former case and $Q=\rho$ in the latter case.  Note that two different types of initial data are recognized, which correspond to two branches in FD as shown in Figure~\ref{fig:PCA4-4} in the case of $\rho \le 1/2$.
%%%%%%%%%%%%%%%%%%%%%%%%%%%%%%%%%%%%%%%%%%%%%%%%%%%%%%%%%%%%%%%%%%%%%
\section{Concluding remarks}  \label{sec:remarks}
%%%%%%%%%%%%%%%%%%%%%%%%%%%%%%%%%%%%%%%%%%%%%%%%%%%%%%%%%%%%%%%%%%%%%
We propose the max-plus expressions to equations and solutions of PCA4-1, 4-2 and 4-3.  We analyze the asymptotic behaviors of solutions and prove the fundamental diagrams mathematically.  The max and min formulas are effectively utilized in the proofs.  In view of the role of basic operations appearing in max-plus algebra, the way of proofs is similar to that for the finite-difference or differential equation.  In this sense, the mathematical technique used is analytic.  It is important to introduce such an analytic technique into the digital systems since it provides us with an alternative approach to study them, and sometimes can lead to rigorous results. \par
  We restrict ourselves to the condition that the value of $u$ is binary in the proof of solutions for PCA4-2 and 4-3. Hence, the max-plus expressions for general solutions proposed only hold under this condition.  On the other hand, solutions for PCA3 and 4-1 are free from this condition.  Their solutions and evaluation of asymptotic behaviors from more general data (real valued data) are valid.  It is a future problem to check the validity of the present expressions for PCA4-2 and 4-3 from more general data.\par
  Finally, we have not yet obtained the max-plus expression for solution to PCA4-4 and PCA of more than four neighbors. The general relation between PCA and its corresponding continuous equation is not clear by using the ultradiscretization method.  It is worthy to solve these problems in the future.
%%%%%%%%%%%%%%%%%%%%%%%%%%%%%%%%%%%%%%%%%%%%%%%%%%%%%%%%%%%%%%%%%%%%%
\appendix
%%%%%%%%%%%%%%%%%%%%%%%%%%%%%%%%%%%%%%%%%%%%%%%%%%%%%%%%%%%%%%%%%%%%%
\section{Formulas about max and min operations}  \label{sec:formulas}
%%%%%%%%%%%%%%%%%%%%%%%%%%%%%%%%%%%%%%%%%%%%%%%%%%%%%%%%%%%%%%%%%%%%%
There are useful formulas to derive or prove relations concerning $\max$ and $\min$ operations. Below we list up such formulas relevant to this paper.\par
  The definition of $\max$ and $\min$ operations are as follows.
\begin{itemize}
\item
$\max(a,b)=\cases{a & if $a\ge b$,\\ b & otherwise}$, \qquad
$\min(a,b)=\cases{a & if $a\le b$,\\ b & otherwise}$.
\end{itemize}
From this definition, $\max$ and $\min$ are related each other.
\begin{itemize}
\item $\max(a,b)=-\min(-a,-b)$,\qquad $\min(a,b)=-\max(-a,-b)$.
\end{itemize}
Most basic formulas satisfy commutative, associative and distributive laws as follows, where $\max$ or $\min$ plays a role of the addition and $+$, i.e., the usual sum of real numbers serves as the multiplication.
\begin{itemize}
\item
  $\max(a,b)=\max(b,a)$,\qquad $\min(a,b)=\min(b,a)$.
\item
  $\max(\max(a,b),c)=\max(a,\max(b,c))$,\qquad $\min(\min(a,b),c)=\min(a,\min(b,c))$.
\item
  $\max(a,b)+c=\max(a+c,b+c)$,\qquad $\min(a,b)+c=\min(a+c,b+c)$.
\end{itemize}
From the commutative and associative laws, we can easily extend
$\max$ and $\min$ operations from binary to multinary.
\begin{itemize}
\item
$\max(a_1,a_2,\ldots)=\hbox{the maximum value among $a_1$, $a_2$, $\ldots$}$,\\
$\min(a_1,a_2,\ldots)=\hbox{the minimum value among $a_1$, $a_2$, $\ldots$}$.
\end{itemize}
We have the following formulas about these multinary operations.
\begin{itemize}
\item
$\max(a_1,a_2,\ldots)=-\min(-a_1,-a_2,\ldots)$,\\
$\min(a_1,a_2,\ldots)=-\max(-a_1,-a_2,\ldots)$.
\item
$\max(\ldots,a_i,\ldots,a_j,\ldots)=\max(\ldots,a_j,\ldots,a_i,\ldots)$,\\
$\min(\ldots,a_i,\ldots,a_j,\ldots)=\min(\ldots,a_j,\ldots,a_i,\ldots)$.
\item
$\max(\max(a_1,a_2,\ldots),b_1,b_2,\ldots)=\max(a_1,a_2,\ldots,b_1,b_2,\ldots)$,\\
$\min(\min(a_1,a_2,\ldots),b_1,b_2,\ldots)=\min(a_1,a_2,\ldots,b_1,b_2,\ldots)$.
\item
$\max(a_1,a_2,\ldots)+b=\max(a_1+b,a_2+b,\ldots)$,\\
$\min(a_1,a_2,\ldots)+b=\min(a_1+b,a_2+b,\ldots)$.
\item
If there exist $i$ such that $a\le b_i$, $\max(a,b_1,b_2,\ldots)=\max(b_1,b_2,\ldots)$.\\
If there exist $i$ such that $a\ge b_i$, $\min(a,b_1,b_2,\ldots)=\min(b_1,b_2,\ldots)$.
\item
$\max(\min(a_1,a_2,\ldots),b_1,b_2,\ldots)=\min(\max(a_1,b_1,b_2,\ldots),\max(a_2,b_1,b_2,\ldots),\ldots)$\\
$\min(\max(a_1,a_2,\ldots),b_1,b_2,\ldots)=\max(\min(a_1,b_1,b_2,\ldots),\min(a_2,b_1,b_2,\ldots),\ldots)$\\
\end{itemize}
The last set of formula regarding the exchange of max and min operations has been widely used in this paper.
%%%%%%%%%%%%%%%%%%%%%%%%%%%%%%%%%%%%%%%%%%%%%%%%%%%%%%%%%%%%%%%%%%%%%
\section{Proof that (\ref{sol PCA4-2}) satisfies (\ref{potential PCA4-2})}  \label{sec:proof solution}
%%%%%%%%%%%%%%%%%%%%%%%%%%%%%%%%%%%%%%%%%%%%%%%%%%%%%%%%%%%%%%%%%%%%%
We prove that (\ref{sol PCA4-2}) satisfies (\ref{potential PCA4-2}) by the mathematical induction.  For $n=1$, it is true since (\ref{sol PCA4-2}) is equivalent to (\ref{potential PCA4-2}) of $n=0$.  Let us assume that (\ref{sol PCA4-2})  satisfies (\ref{potential PCA4-2}) for $n=m$  ($m\ge1$).  Then $f_j^{m+1}$ is expressed by $f_j^1$ as
\begin{equation*}
  f_j^{m+1}=\max_{0\le k\le m}\Bigl(\min\Bigl(f_{j-2m+3k}^1-\frac{k}{2},\min_{k+1\le i\le m}f_{j-2m+3i}^1+\frac{i}{2}\Bigr)\Bigr).
\end{equation*}
Using (\ref{potential PCA4-2}), we obtain
\begin{equation*}
\fl\eqalign{f_j^{m+1} &= \max_{0\le k\le m}\Bigl(\min\Bigl( \\
&\qquad \max\Bigl(\min\Bigl(f_{j-2m+3k-2}^0,f_{j-2m+3k+1}^0+\frac{1}{2}\Bigr),f_{j-2m+3k+1}^0-\frac{1}{2}\Bigr)-\frac{k}{2}, \\
&\qquad \min_{k+1\le i\le m}\Bigl(\max\Bigl(\min\Bigl(f_{j-2m+3i-2}^0,f_{j-2m+3i+1}^0+\frac{1}{2}\Bigr), f_{j-2m+3i+1}^0-\frac{1}{2}\Bigr)\Bigr)+\frac{i}{2} \\
&\quad \Bigr)\Bigr).}
\end{equation*}
Using the max-min formula, we have
\begin{equation*}
\fl\eqalign{f_j^{m+1} &= \max_{0\le k\le m}\Bigl(\min\Bigl( \\
&\qquad \min\Bigl(\max\Bigl(f_{j-2m+3k-2}^0,f_{j-2m+3k+1}^0-\frac{1}{2}\Bigr),\\
&\qquad\qquad \max\Bigl(f_{j-2m+3k+1}^0+\frac{1}{2},f_{j-2m+3k+1}^0-\frac{1}{2}\Bigr)\Bigr)-\frac{k}{2}, \\
&\qquad \min_{k+1\le i\le m}\Bigl(\min\Bigl(\max\Bigl(f_{j-2m+3i-2}^0,f_{j-2m+3i+1}^0-\frac{1}{2}\Bigr), \\
&\qquad\qquad\qquad\qquad \max\Bigl(f_{j-2m+3i+1}^0+\frac{1}{2},f_{j-2m+3i+1}^0-\frac{1}{2}\Bigr)\Bigr)\Bigr)+\frac{i}{2} \\
&\quad \Bigr)\Bigr) \\
&= \max_{0\le k\le m}\Bigl(\min\Bigl( \\
&\qquad \max\Bigl(f_{j-2m+3k-2}^0-\frac{k}{2},f_{j-2m+3k+1}^0-\frac{k+1}{2}\Bigr),f_{j-2m+3k+1}^0-\frac{k-1}{2}, \\
&\qquad \min_{k+1\le i\le m}\Bigl(\max\Bigl(f_{j-2m+3i-2}^0+\frac{i}{2},f_{j-2m+3i+1}^0+\frac{i-1}{2}\Bigr), f_{j-2m+3i+1}^0+\frac{i+1}{2}\Bigr) \\
&\quad \Bigr)\Bigr). \\
}
\end{equation*}
Expanding the terms, we have
\begin{equation*}
\fl\eqalign{
f_j^{m+1}=&\max_{0\le k\le m}\Bigl(\min\Bigl( \\
&\quad\max\Bigl(f_{j-2m+3k-2}^0-\frac{k}{2},f_{j-2m+3k+1}^0-\frac{k+1}{2}\Bigr),f_{j-2m+3k+1}^0-\frac{k-1}{2}, \\
&\quad\underline{\max\Bigl(f_{j-2m+3k+1}^0+\frac{k+1}{2},f_{j-2m+3k+4}^0+\frac{k}{2}\Bigr)},f_{j-2m+3k+4}^0+\frac{k+2}{2}, \\
&\quad\underline{\max\Bigl(f_{j-2m+3k+4}^0+\frac{k+2}{2},f_{j-2m+3k+7}^0+\frac{k+1}{2}\Bigr)},f_{j-2m+3k+7}^0+\frac{k+3}{2}, \\
& \ldots, \\
&\quad\underline{\max\Bigl(f_{j+m-2}^0+\frac{m}{2},f_{j+m+1}^0+\frac{m-1}{2}\Bigr)},f_{j+m+1}^0+\frac{m+1}{2}\\
&\Bigr)\Bigr).
}
\end{equation*}
Since the underlined terms are negligible, we obtain
\begin{equation*}
\fl\eqalign{
f_j^{m+1}=\max_{0\le k\le m}\Bigl(&\min\Bigl(
\max\Bigl(f_{j-2m+3k-2}^0-\frac{k}{2},f_{j-2m+3k+1}^0-\frac{k+1}{2}\Bigr), \\
& f_{j-2m+3k+1}^0-\frac{k-1}{2}, f_{j-2m+3k+4}^0+\frac{k+2}{2}, f_{j-2m+3k+7}^0+\frac{k+3}{2}, \\
& \ldots,f_{j+m+1}^0+\frac{m+1}{2}\Bigr)\Bigr).
}
\end{equation*}
Using the max-min formula again, we have
\begin{equation*}
\fl\eqalign{
f_j^{m+1}=\max_{0\le k\le m}\Bigl(&\min\Bigl(f_{j-2m+3k-2}^0-\frac{k}{2},f_{j-2m+3k+1}^0-\frac{k-1}{2}, \\
&\qquad f_{j-2m+3k+4}^0+\frac{k+2}{2}, f_{j-2m+3k+7}^0+\frac{k+3}{2}, \ldots, f_{j+m+1}^0+\frac{m+1}{2}\Bigr),\\
&\min\Bigl(f_{j-2m+3k+1}^0-\frac{k+1}{2}, \\
&\qquad f_{j-2m+3k+4}^0+\frac{k+2}{2}, f_{j-2m+3k+7}^0+\frac{k+3}{2}, \ldots, f_{j+m+1}^0+\frac{m+1}{2}\Bigr)\Bigr).
}
\end{equation*}
Expanding the terms, we have
\begin{equation*}
\fl\eqalign{
f_j^{m+1}=\max\Bigl(&\min\Bigl(f_{j-2m-2}^0,f_{j-2m+1}^0+\frac{1}{2}, f_{j-2m+4}^0+1, \ldots, f_{j+m+1}^0+\frac{m+1}{2}\Bigr),\\
&\min\Bigl(f_{j-2m+1}^0-\frac{1}{2}, f_{j-2m+4}^0+1, f_{j-2m+7}^0+\frac{3}{2}, \ldots, f_{j+m+1}^0+\frac{m+1}{2}\Bigr), \\
&\underline{\min\Bigl(f_{j-2m+1}^0-\frac{1}{2},f_{j-2m+4}^0, f_{j-2m+7}^0+\frac{3}{2}, \ldots, f_{j+m+1}^0+\frac{m+1}{2}\Bigr)},\\
&\min\Bigl(f_{j-2m+4}^0-1, f_{j-2m+7}^0+\frac{3}{2}, f_{j-2m+10}^0+2, \ldots, f_{j+m+1}^0+\frac{m+1}{2}\Bigr), \\
&\underline{\min\Bigl(f_{j-2m+4}^0-1,f_{j-2m+7}^0-\frac{1}{2}, f_{j-2m+10}^0+2, \ldots, f_{j+m+1}^0+\frac{m+1}{2}\Bigr)},\\
&\min\Bigl(f_{j-2m+7}^0-\frac{3}{2}, f_{j-2m+10}^0+2,f_{j-2m+13}^0+\frac{5}{2}, \ldots, f_{j+m+1}^0+\frac{m+1}{2}\Bigr), \\
&\ldots, \\
&\underline{\min\Bigl(f_{j+m-5}^0-\frac{m-1}{2},f_{j+m-2}^0-\frac{m-2}{2}, f_{j+m+1}^0+\frac{m+1}{2}\Bigr)}, \\
&\min\Bigl(f_{j+m-2}^0-\frac{m}{2}, f_{j+m+1}^0+\frac{m+1}{2}\Bigr), \\
&\underline{\min\Bigl(f_{j+m-2}^0-\frac{m}{2},f_{j+m+1}^0-\frac{m-1}{2}, f_{j+m+1}^0+\frac{m+1}{2}\Bigr)}, \\
&\min\Bigl(f_{j+m+1}^0-\frac{m+1}{2}, \underline{f_{j+m+1}^0+\frac{m+1}{2}}\Bigr)\Bigr).
}
\end{equation*}
Since the underlined terms are negligible, we obtain
\begin{equation*}
\fl\eqalign{
f_j^{m+1}=\max\Bigl(&\min\Bigl(f_{j-2m-2}^0,f_{j-2m+1}^0+\frac{1}{2}, f_{j-2m+4}^0+1, \ldots, f_{j+m+1}^0+\frac{m+1}{2}\Bigr),\\
&\min\Bigl(f_{j-2m+1}^0-\frac{1}{2}, f_{j-2m+4}^0+1, f_{j-2m+7}^0+\frac{3}{2}, \ldots, f_{j+m+1}^0+\frac{m+1}{2}\Bigr), \\
&\min\Bigl(f_{j-2m+4}^0-1, f_{j-2m+7}^0+\frac{3}{2}, f_{j-2m+10}^0+2, \ldots, f_{j+m+1}^0+\frac{m+1}{2}\Bigr), \\
&\ldots, \\
&\min\Bigl(f_{j+m-2}^0-\frac{m}{2}, f_{j+m+1}^0+\frac{m+1}{2}\Bigr),\\
&f_{j+m+1}^0-\frac{m+1}{2}\Bigr)\Bigr) \\
=\max_{0\le k\le m+1}\Bigl(&\min\Bigl(f_{j-2m+3k-2}^0-\frac{k}{2},
\min_{k+1\le i\le m+1}f_{j-2m+3i-2}^0+\frac{i}{2}\Bigr)\Bigr).
}
\end{equation*}
The above equation is nothing but (\ref{sol PCA4-2}) for $n=m+1$.  Thus the proof is complete by the mathematical induction.
%%%%%%%%%%%%%%%%%%%%%%%%%%%%%%%%%%%%%%%%%%%%%%%%%%%%%%%%%%%%%%%%%%%%%
\section{Proof about relations from (\ref{asymp PCA4-2 f}) to (\ref{asymp PCA4-2 Q})}  \label{sec:proof asymptotic}
%%%%%%%%%%%%%%%%%%%%%%%%%%%%%%%%%%%%%%%%%%%%%%%%%%%%%%%%%%%%%%%%%%%%%
The asymptotic behavior of solutions to PCA4-2 is different for three ranges of $\rho$, $0\le\rho\le1/3$, $1/3\le\rho\le2/3$ and $2/3\le\rho\le1$. Therefore we consider three cases separately. First, we expand the solution (\ref{sol PCA4-2}) for $n\gg0$ as follows to make evaluation easier;
\begin{equation}  \label{expanded sol PCA4-2}
\fl\eqalign{
  f_j^n=\max\Bigl(
&\min\Bigl(f_{j-2n}^0,f_{j-2n+3}^0+\frac{1}{2},f_{j-2n+6}^0+1,\ldots,f_{j+n}^0+\frac{n}{2}\Bigr),\\
&\min\Bigl(f_{j-2n+3}^0-\frac{1}{2},f_{j-2n+6}^0+1,f_{j-2n+9}^0+\frac{3}{2},\ldots,f_{j+n}^0+\frac{n}{2}\Bigr),\\
&\min\Bigl(f_{j-2n+6}^0-1,f_{j-2n+9}^0+\frac{3}{2},f_{j-2n+12}^0+2,\ldots,f_{j+n}^0+\frac{n}{2}\Bigr),\\
&\ldots,\\
&\min\Bigl(f_{j+n-3}^0-\frac{n-1}{2},f_{j+n}^0+\frac{n}{2}\Bigr),\\
&f_{j+n}^0-\frac{n}{2}\Bigr).
}
\end{equation}
%%%%%%%%%%%%%%%%%%%%%%%%%%%%%%%%%%%%%%%%%%%%%%%%%%%%%%%%%%%%%%%%%%%%%
\subsection{Case of $\mathit{0\le\rho\le1/3}$}
%%%%%%%%%%%%%%%%%%%%%%%%%%%%%%%%%%%%%%%%%%%%%%%%%%%%%%%%%%%%%%%%%%%%%
In this case, we have
\begin{equation*}
  f_{j+3pK}-f_j^n+\frac{pK}{2}=\sum_{k=j+1}^{j+3pK}\Bigl(u_k^n-\frac{1}{2}\Bigr)+\frac{pK}{2}=3pK\Bigl(\rho-\frac{1}{3}\Bigr)\le0,
\end{equation*}
for any $j$ and $n$ where $p$ is a non-negative integer and $K$ is the spatial
period.  Therefore we have
\begin{equation}  \label{ineq PCA4-2}
  f_j^n\ge f_{j+3pK}^n+\frac{pK}{2}.
\end{equation}
Let us define $g_j^n$ by
\begin{equation*}
\fl\eqalign{
g_j^n &= \min_{0\le k\le K-1}\Bigl(f_{j+n-3k}^0+\frac{n-k}{2}\Bigr) \\
&= \min\Bigl(f_{j+n-3K+3}^0+\frac{n-K+1}{2}, f_{j+n-3K+6}^0+\frac{n-K+2}{2}, \ldots, f_{j+n}^0+\frac{n}{2}\Bigr).
}
\end{equation*}
For $p>0$ and $0\le q<K$,
\begin{equation*}
  f_{j+n-3(pK+q)}^0+\frac{n-pK-q}{2} \ge f_{j+n-3q}^0+\frac{n-q}{2}
\end{equation*}
holds from (\ref{ineq PCA4-2}). Therefore the first $\min$ term of (\ref{expanded sol PCA4-2}) is equal to $g_j^n$ for $n\gg0$.  We can make a similar evaluation from the second to the $(n-K+1)$-th $\min$ terms of (\ref{expanded sol PCA4-2}) as
\begin{equation*}
\fl\eqalign{
\min\Bigl(f_{j-2n+3}^0-\frac{1}{2},f_{j-2n+6}^0+1,f_{j-2n+9}^0+\frac{3}{2},\ldots,f_{j+n}^0+\frac{n}{2}\Bigr) = \min\Bigl(f_{j-n+3}^0-\frac{1}{2},g_j^n\Bigr), \\
\min\Bigl(f_{j-2n+6}^0-1,f_{j-2n+9}^0+\frac{3}{2},f_{j-2n+12}^0+2,\ldots,f_{j+n}^0+\frac{n}{2}\Bigr) = \min\Bigl(f_{j-2n+6}^0-1,g_j^n\Bigr), \\
\ldots,\\
\min\Bigl(f_{j+n-3K}^0-\frac{n-K}{2},f_{j+n-3K+3}^0+\frac{n-K+1}{2},f_{j+n-3K+6}^0+\frac{n-K+2}{2},\\
\qquad\ldots,f_{j+n}^0+\frac{n}{2}\Bigr)=\min\Bigl(f_{j+n-3K}^0-\frac{n-K}{2},g_j^n\Bigr).
}
\end{equation*}
Moreover, the relation
\begin{equation*}
  f_j^n-f_{j-1}^n=u_j^n-\frac{1}{2}=\pm\frac{1}{2}
\end{equation*}
holds since $u_j^n=0$ or 1.  It implies that the difference between
$f_j^n$ and $f_{j\pm p}^n$ is finite if $p$ is finite.  Therefore we
can evaluate from the $(n-K+2)$-th to the penultimate terms
of (\ref{expanded sol PCA4-2}) for $n\gg0$ as
\begin{equation*}
\fl\eqalign{
\min\Bigl(f_{j+n-3K+3}^0-\frac{n-K+1}{2},f_{j+n-3K+6}^0+\frac{n-K+2}{2},\ldots,f_{j+n}^0+\frac{n}{2}\Bigr) \\
\qquad = f_{j+n-3K+3}^0-\frac{n-K+1}{2}, \\
\min\Bigl(f_{j+n-3K+6}^0-\frac{n-K+2}{2},f_{j+n-3K+9}^0+\frac{n-K+3}{2},\ldots,f_{j+n}^0+\frac{n}{2}\Bigr) \\
\qquad = f_{j+n-3K+6}^0-\frac{n-K+2}{2}, \\
\ldots,\\
\min\Bigl(f_{j+n-6}^0-\frac{n-2}{2},f_{j+n-3}^0+\frac{n-1}{2},f_{j+n}^0+\frac{n}{2}\Bigr)=f_{j+n-6}^0-\frac{n-2}{2}, \\
\min\Bigl(f_{j+n-3}^0-\frac{n-1}{2},f_{j+n}^0+\frac{n}{2}\Bigr)=f_{j+n-3}^0-\frac{n-1}{2}.
}
\end{equation*}
  Thus (\ref{expanded sol PCA4-2}) becomes
\begin{equation*}
\fl\eqalign{
  f_j^n&=\max\Bigl(g_j^n,\min\Bigl(f_{j-n+3}^0-\frac{1}{2},g_j^n\Bigr),\ldots,\min\Bigl(f_{j+n-3K}^0-\frac{n-K}{2},g_j^n\Bigr),\\
  &\qquad\qquad f_{j+n-3K+3}^0-\frac{n-K+1}{2},\ldots,f_{j+n}^0-\frac{n}{2}\Bigr) \\
 &= g_j^n.
}
\end{equation*}
It coincides with (\ref{asymp PCA4-2 f}) in the case of
$0\le\rho\le1/3$.\par
  Since we can easily show
\begin{equation*}
  g_j^{n+1}=g_{j+1}^n+\frac{1}{2},
\end{equation*}
we have
\begin{equation*}
  u_j^{n+1}=g_j^{n+1}-g_{j-1}^{n+1}+\frac{1}{2}=g_{j+1}^n-g_j^n+\frac{1}{2}=u_{j+1}^n.
\end{equation*}
Substituting $u_j^{n+1}=u_{j+1}^n$ into the evolution equation (\ref{eq PCA4-2}), we obtain
\begin{equation*}
  u_{j+1}^n+q(u_{j-1}^n,u_j^n,u_{j+1}^n)=u_j^n+q(u_{j-2}^n,u_{j-1}^n,u_j^n).
\end{equation*}
Therefore $u_j^n+q(u_{j-2}^n,u_{j-1}^n,u_j^n)$ is a constant irrespective of $j$.  Since there exists at least a site $j$ such that $(u_{j-2}^n,u_{j-1}^n,u_j^n)=(0,0,0)$ or $(1,0,0)$ for $0\le\rho\le1/3$, we obtain $u_j^n+q(u_{j-2}^n,u_{j-1}^n,u_j^n)=0$ for any $j$.  Thus we have
\begin{equation*}
  Q=\frac{1}{K}\sum_{j=1}^K q(u_{j-2}^n,u_{j-1}^n,u_j^n)=-\frac{1}{K}\sum_{j=1}^K u_j^n=-\rho.
\end{equation*}
The above results coincide with (\ref{asymp PCA4-2 u}), (\ref{asymp PCA4-2 q}) and (\ref{asymp PCA4-2 Q}) in the case of $0\le\rho\le1/3$.
%%%%%%%%%%%%%%%%%%%%%%%%%%%%%%%%%%%%%%%%%%%%%%%%%%%%%%%%%%%%%%%%%%%%%
\subsection{Case of $\mathit{1/3\le\rho\le2/3}$}
%%%%%%%%%%%%%%%%%%%%%%%%%%%%%%%%%%%%%%%%%%%%%%%%%%%%%%%%%%%%%%%%%%%%%
In this case, the inequality
\begin{equation}  \label{cond mid rho}
  f_{j+3pK}^n-\frac{pK}{2}\le f_j^n\le f_{j+3pK}^n+\frac{pK}{2},
\end{equation}
holds for any $j$ and $n$ where $p>0$.  Under this condition, the first to the $(K-1)$-th $\min$ terms of (\ref{expanded sol PCA4-2}) for $n\gg0$ are evaluated as follows:
\begin{equation*}
\fl\eqalign{
\min\Bigl(f_{j-2n}^0,f_{j-2n+3}^0+\frac{1}{2},\ldots,f_{j+n}^0+\frac{n}{2}\Bigr) \\
\qquad = \min\Bigl(f_{j-2n}^0,f_{j-2n+3}^0+\frac{1}{2},\ldots,f_{j-2n+3(K-1)}^0+\frac{K-1}{2}\Bigr), \\
\min\Bigl(f_{j-2n+3}^0-\frac{1}{2},f_{j-2n+6}^0+1,\ldots,f_{j+n}^0+\frac{n}{2}\Bigr) \\
\qquad = \min\Bigl(f_{j-2n+3}^0-\frac{1}{2},f_{j-2n+6}^0+1,\ldots,f_{j-2n+3K}^0+\frac{K}{2}\Bigr), \\
\ldots, \\
\min\Bigl(f_{j-2n+3(K-2)}^0-\frac{K-2}{2},f_{j-2n+3(K-1)}^0+\frac{K-1}{2},\ldots,f_{j+n}^0+\frac{n}{2}\Bigr) \\
\qquad =\min\Bigl(f_{j-2n+3(K-2)}^0-\frac{K-2}{2},f_{j-2n+3(K-1)}^0+\frac{K-1}{2},\\
\qquad\qquad\qquad\ldots,f_{j-2n+3(2K-3)}^0+\frac{2K-3}{2}\Bigr).
}
\end{equation*}
Regarding the rest $\min$ terms of (\ref{expanded sol PCA4-2}), they are expressed generally by
\begin{equation*}
  r=\min\Bigl(f_{j-2n+3k}^0-\frac{k}{2},\min_{k+1\le i\le n}f_{j-2n+3i}^0+\frac{i}{2}\Bigr),
\end{equation*}
where $K-1\le k\le n$.  We can evaluate them similarly as above and obtain
\begin{equation*}
r_k = \min\Bigl(f_{j-2n+3k}^0-\frac{k}{2},\min_{k+1\le i\le \min(n,k+K-1)}f_{j-2n+3i}^0+\frac{i}{2}\Bigr).
\end{equation*}
Then the inequality
\begin{equation*}
\fl\eqalign{
\Bigl(f_{j-2n+3i}^0+\frac{i}{2}\Bigr)-\Bigl(f_{j-2n+3k}^0-\frac{k}{2}\Bigr)
= \sum_{l=j-2n+3k+1}^{j-2n+3i}\Bigl(u_l^0-\frac{1}{2}\Bigr)+\frac{i+k}{2} \\
\ge-\frac{3(i-k)}{2}+\frac{i+k}{2}=2k-i \ge 2k-\min(n,k+K-1) \\
=\max(2k-n,k-K+1)\ge0
}
\end{equation*}
holds since $i\le\min(n,k+K-1)$ and $k\ge K-1$.  Thus we obtain
\begin{equation*}
  r_k=f_{j-2n+3k}-\frac{k}{2}.
\end{equation*}
Moreover, note that
\begin{equation*}
  r_{k+pK}\le r_k
\end{equation*}
holds for $p>0$ by (\ref{cond mid rho}).\par

Therefore we obtain the final results about the asymptotic evaluation of $f_j^n$ for $n\gg0$ in the case of $1/3\le\rho\le2/3$ as
\begin{equation*}
\fl\eqalign{
  f_j^n=\max\Bigl(
&\min\Bigl(f_{j-2n}^0,f_{j-2n+3}^0+\frac{1}{2},\ldots,f_{j-2n+3(K-1)}^0+\frac{K-1}{2}\Bigr), \\
&\min\Bigl(f_{j-2n+3}^0-\frac{1}{2},f_{j-2n+6}^0+1,\ldots,f_{j-2n+3K}^0+\frac{K}{2}\Bigr), \\
&\ldots, \\
&\min\Bigl(f_{j-2n+3(K-2)}^0-\frac{K-2}{2},f_{j-2n+3(K-1)}^0+\frac{K-1}{2},\\
&\qquad\ldots,f_{j-2n+3(2K-3)}^0+\frac{2K-3}{2}\Bigr), \\
&f_{j-2n+3(K-1)}^0-\frac{K-1}{2}, f_{j-2n+3K}^0-\frac{K}{2}, \ldots, f_{j-2n+3(2K-2)}^0-(K-1)\Bigr).
}
\end{equation*}
It coincides with (\ref{asymp PCA4-2 f}) in the case of $1/3\le\rho\le2/3$.\par
  It can be easily shown that the relation $f_j^{n+1}=f_{j-2}^n$ holds for $f_j^n$ in the above equation. Hence so does the relation $u_j^{n+1}=u_{j-2}^n$.  Substituting it into (\ref{eq PCA4-2}), we obtain
\begin{equation*}
  u_{j-2}^n+u_{j-1}^n-q(u_{j-2}^n,u_{j-1}^n,u_j^n)=u_{j-1}^n+u_j^n-q(u_{j-1}^n,u_j^n,u_{j+1}^n),
\end{equation*}
and $u_{j-2}^n+u_{j-1}^n-q(u_{j-2}^n,u_{j-1}^n,u_j^n)$ is a constant irrespective of $j$.  Since there exists at least a site $j$ such that $(u_{j-2}^n,u_{j-1}^n,u_j^n)=(1,0,0)$ or $(1,1,0)$ for $1/3\le\rho\le2/3$, we have $u_{j-2}^n+u_{j-1}^n-q(u_{j-2}^n,u_{j-1}^n,u_j^n)=1$.  Thus we derive
\begin{equation*}
  Q=\frac{1}{K}\sum_{j=1}^K q(u_{j-2}^n,u_{j-1}^n,u_j^n)=\frac{1}{K}\sum_{j=1}^K(u_{j-2}^n+u_{j-1}^n-1)=2\rho-1.
\end{equation*}
The above results coincide with (\ref{asymp PCA4-2 u}), (\ref{asymp PCA4-2 q}) and (\ref{asymp PCA4-2 Q}) in the case of $1/3\le\rho\le2/3$.
%%%%%%%%%%%%%%%%%%%%%%%%%%%%%%%%%%%%%%%%%%%%%%%%%%%%%%%%%%%%%%%%%%%%%
\subsection{Case of $\mathit{2/3\le\rho\le1}$}
%%%%%%%%%%%%%%%%%%%%%%%%%%%%%%%%%%%%%%%%%%%%%%%%%%%%%%%%%%%%%%%%%%%%%
Substituting the transformation $u=1-\tilde u$ into (\ref{eq PCA4-2}), we obtain
\begin{equation*}
\cases{
  \tilde u_j^{n+1}=\tilde u_j^n+\tilde q(\tilde u_{j-2}^n,\tilde u_{j-1}^n,\tilde u_j^n)-\tilde q(\tilde u_{j-1}^n,\tilde u_j^n,\tilde u_{j+1}^n),\\
  \tilde q(a,b,c)=-\min(\max(c-1,1-a-b), c)
}.
\end{equation*}
Using the max-min formulas, we have
\begin{equation*}
\fl\eqalign{
  \tilde q(a,b,c)&=\max(\min(1-c,a+b-1), -c)=\min(\max(1-c,-c),\max(a+b-1,-c)) \\
 &=\min(\max(-c,a+b-1),1-c).
}
\end{equation*}
Since $\tilde q$ is the same as $q$, the evolution equation (\ref{eq PCA4-2}) is self-conjugate, that is, it is invariant under the above transformation.  Therefore, the behaviors for solutions of $0\le\rho\le1/3$ and $2/3\le\rho\le1$ are symmetric to each other and there exist one-to-one correspondence between the solutions.  Since $u_j^{n+1}=u_{j+1}^n$ holds for $n\gg0$ in the case of $0\le\rho\le1/3$, $1-u_j^{n+1}=1-u_{j+1}^n$, that is, $u_j^{n+1}=u_{j+1}^n$ again holds in the case of $2/3\le\rho\le1$.  Substituting this relation to (\ref{eq PCA4-2}), $u_j^n+q(u_{j-2}^n,u_{j-1}^n,u_j^n)$ is again a constant. Since there exists at least a site $j$ such that $(u_{j-2}^n,u_{j-1}^n,u_j^n)=(1,1,0)$ or $(1,1,1)$ for $2/3\le\rho\le1$, we obtain $u_j^n+q(u_{j-2}^n,u_{j-1}^n,u_j^n)=1$ for any $j$.  Thus we have
\begin{equation}
  Q=\frac{1}{K}\sum_{j=1}^K q(u_{j-2}^n,u_{j-1}^n,u_j^n)=\frac{1}{K}\sum_{j=1}^K (1-u_j^n)=1-\rho.
\end{equation}
These results coincide with (\ref{asymp PCA4-2 u}), (\ref{asymp
PCA4-2 q}) and (\ref{asymp PCA4-2 Q}) in the case of
$2/3\le\rho\le1$.\par
  Moreover, introducing the transformation $f_j^n=-\tilde f_j^n$ with $u_j^n=1-\tilde u_j^n$, we obtain
\begin{equation*}
  \tilde u_j^n=\tilde f_j^n-\tilde f_{j-1}^n+\frac{1}{2}
\end{equation*}
from (\ref{trans PCA4-2}), and
\begin{equation*}
\eqalign{
  \tilde f_j^{n+1}&=\min\Bigl(\max\Bigl(\tilde f_{j-2}^n,\tilde f_{j+1}^n-\frac{1}{2}\Bigr),\tilde f_{j+1}^n+\frac{1}{2}\Bigr) \\
  &=\max\Bigl(\min\Bigl(\tilde f_{j-2}^n,\tilde f_{j+1}^n+\frac{1}{2}\Bigr),\tilde f_{j-1}^n-\frac{1}{2}\Bigr)
}
\end{equation*}
from (\ref{potential PCA4-2}).  These relations are in the same form of (\ref{trans PCA4-2}) and (\ref{potential PCA4-2}).  Thus the asymptotic solution $f_j^n$ for $2/3\le\rho\le1$ is derived from that for $0\le\rho\le1/3$ as
\begin{equation*}
\fl  f_j^n=-\min_{0\le k\le K-1}\Bigl(-f_{j+n-3k}^0+\frac{n-k}{2}\Bigr)=\max_{0\le k\le K-1}\Bigl(f_{j+n-3k}^0-\frac{n-k}{2}\Bigr)
\end{equation*}
It coincides with (\ref{asymp PCA4-2 f}) in the case of $2/3\le\rho\le1$.\par
  Note that the above symmetry of bit inversion holds for any $n$ as well as for $n\gg0$ and it gives another expression of solution (\ref{sol PCA4-2}) as follows:
\begin{equation*}
\eqalign{
  f_j^n&=-\max_{0\le k\le n}\Bigl(\min\Bigl(-f_{j-2n+3k}^0-\frac{k}{2},\min_{k+1\le i\le n}-f_{j-2n+3i}^0+\frac{i}{2}\Bigr)\Bigr) \\
  &=\min_{0\le k\le n}\Bigl(\max\Bigl(f_{j-2n+3k}^0+\frac{k}{2},\max_{k+1\le i\le n}f_{j-2n+3i}^0-\frac{i}{2}\Bigr)\Bigr).
}
\end{equation*}
%%%%%%%%%%%%%%%%%%%%%%%%%%%%%%%%%%%%%%%%%%%%%%%%%%%%%%%%%%%%%%%%%%%%
\section*{References}
%%%%%%%%%%%%%%%%%%%%%%%%%%%%%%%%%%%%%%%%%%%%%%%%%%%%%%%%%%%%%%%%%%%%%

\end{document}